\documentclass[twocolumn,3p,authoryear]{elsarticle}

\usepackage{natbib}
\bibpunct{(}{)}{;}{a}{}{,}
\usepackage{color}
\definecolor{darkgreen}{RGB}{0,142,128}
\definecolor{mygreen}{RGB}{0,200,50}

\usepackage{hyperref}
\hypersetup{colorlinks,citecolor=blue,linkcolor=blue}

\journal{Advances in Space Research}

\usepackage{amsmath}
\usepackage{amsthm}
\usepackage{amsfonts}
\usepackage{amssymb}
\usepackage{url}
\usepackage{subfigure}
\usepackage{multirow}

\newcommand{\U}{\mathbf{u}}

\newcommand{\dr}{\partial_r}
\newcommand{\drr}{\partial^2_{rr}}
\newcommand{\dth}{\partial_\theta}
\newcommand{\dphi}{\partial_\varphi}

\newcommand{\bnab}{\nabla}
\newcommand{\er}{\mathbf{e}_r}
\newcommand{\etheta}{\mathbf{e}_\theta}
\newcommand{\ephi}{\mathbf{e}_\varphi}
\newcommand{\rot}{\bnab\times}
\newcommand{\Div}{\bnab\cdot}

\newcommand{\grad}{\bnab}

\newcommand{\dint}[1]{{\rm d}{#1}}

\newcommand{\Rlm}[1]{\mathbf{R}^{m_{#1}}_{l_{#1}}}
\newcommand{\Slm}[1]{\mathbf{S}^{m_{#1}}_{l_{#1}}}
\newcommand{\Tlm}[1]{\mathbf{T}^{m_{#1}}_{l_{#1}}}

\newcommand{\MAlm}[1]{\mathcal{A}^{l_{#1}}_{m_{#1}}}
\newcommand{\MBlm}[1]{\mathcal{B}^{l_{#1}}_{m_{#1}}}
\newcommand{\MClm}[1]{\mathcal{C}^{l_{#1}}_{m_{#1}}}

\newcommand{\Ylm}[1]{Y_{l_{#1}}^{m_{#1}}}
\newcommand{\Ylmn}[1]{\mathbf{Y}_{l_{#1},l_{#1}+\nu_{#1}}^{\hspace{0.1cm}m_{#1}}}
\newcommand{\Xlmn}[1]{X_{#1;l_{#1},l_{#1}+\nu_{#1}}^{\hspace{0.1cm}m_{#1}}}

\newcommand{\sumlm}[1]{\sum_{l_{#1},m_{#1}}}
\newcommand{\sumlmn}[1]{\sum_{l_{#1}=0}^{\infty}\sum_{m_{#1} = - l_{#1}}^{l_{#1}}\sum_{\nu_{#1}=-1}^1}

\newcommand{\mysum}[3]{\sum_{\substack{l_{#1},l_{#2} = 0 \\ l_{#3} \ge |l_{#1}-l_{#2}| \\
      l_{#3}\le l_{#1} + l_{#2}}  }^{\infty}
      \sum_{\substack{m_{#1} = -l_{#1} \\m_{#2} = -l_{#2} \\ m_{#1}+m_{#2}=m_{#3}}
}^{l_{#1},l_{#2}}\sum_{\nu_{#1},\nu_{#2}} }

\newcommand{\reff}[1]{{#1}}
\newcommand{\refff}[1]{{#1}}

\newcommand{\aap}{\textit{A\&A} }
\newcommand{\apj}{\textit{ApJ} }
\newcommand{\apjs}{\textit{ApJ Supp. Series} }
\newcommand{\apjl}{\textit{ApJL} }
\newcommand{\lrsp}{\textit{LRSP} }
\newcommand{\gafd}{\textit{Geophys. Astrophys. Fluid Dyn.} }
\newcommand{\mnras}{\textit{MNRAS} }
\newcommand{\jgr}{\textit{J. of Geophys. Res.} }
\newcommand{\jfm}{\textit{J. of Fluid mech.} }
\newcommand{\solphys}{\textit{Sol. Phys.} }
\newcommand{\ssr}{\textit{Space Sci Rev} }

\begin{document}

\begin{frontmatter}

\title{Modelling turbulent stellar convection zones: sub-grid scales
  effects}

\author[udm,cea]{A. Strugarek\corref{cor1}}
\ead{strugarek@astro.umontreal.ca}
\author[udm]{P. Beaudoin}
\author[cea]{A. S. Brun}
\author[udm]{P. Charbonneau}
\author[cea]{S. Mathis}
\author[reading]{P. K. Smolarkiewicz}

\cortext[cor1]{Corresponding author}

\address[udm]{D\'epartement de physique, Universit\'e de Montr\'eal, C.P. 6128 Succ. Centre-Ville, Montr\'eal, QC H3C-3J7, Canada}
\address[cea]{Laboratoire AIM Paris-Saclay, CEA/DSM Universit\'e
  Paris-Diderot CNRS, IFRU/SAp, F-91191 Gif-sur-Yvette, France.}
\address[reading]{European Centre for Medium-Range Weather Forecasts, Reading RG2 9AX, UK}

\begin{abstract}
The impressive development of global numerical simulations of
turbulent stellar
interiors unveiled a variety of possible differential rotation (solar
or anti-solar), meridional circulation (single or multi-cellular), and
dynamo states (stable large scale toroidal field or periodically
reversing magnetic fields).
Various numerical schemes, based on the so-called
anelastic set of equations, were used to obtain these results. It
appears today mandatory to assess their robustness
with respect to the details of the numerics, and in
particular to the treatment of turbulent sub-grid scales. We report on
an ongoing comparison between two global models, the ASH and
EULAG codes. In EULAG the sub-grid scales are treated implicitly by the
numerical scheme, while in ASH their effect is generally modelled by using
enhanced dissipation coefficients. We characterize the sub-grid scales
effect in a turbulent convection simulation with EULAG. We assess
their effect at each resolved scale with a detailed energy budget. We
derive equivalent eddy-diffusion coefficients and use the
derived diffusivities in twin ASH numerical simulations. We find a
good agreement between the large-scale flows developing in the two
codes in the hydrodynamic regime, which
encourages further
investigation in the magnetohydrodynamic regime for various dynamo solutions.
\end{abstract}


\end{frontmatter}

\section{Introduction}
\label{sec:introduction}

Cool stars are known to possess a substantial convection zone
forming their outer layer. The convective motions participate in the
self-organization of the interior of the star and in particular in
the sustainment of a large scale differential rotation
\citep[\textit{e.g.}][and references
therein]{Brun:2002gi,Featherstone:2015bv}, and a
potentially cyclic magnetism
\citep[\textit{e.g.}][and references therein]{Ghizaru:2010im,Kapyla:2012dg,Augustson:2015er,Brun:2015kc}. Understanding the properties of
solar and stellar convection is hence of prior importance to understand
the joint evolution of the stellar rotation rate and large-scale
magnetic fields along the Main Sequence. 

Thanks to helioseismology
\citep{ChristensenDalsgaard:1991iv,Basu:1997wd}, we know that the
solar stratification is close to \reff{adiabatic} down to 0.71 solar radii,
leaving little doubt that convective motions extend
down to such depths. However, the amplitude of these deep convective flows
in the Sun is today the subject of a stimulating controversy. Using
time-distance helioseismology,
\citet{Hanasoge:2010jd,Hanasoge:2012fr} obtained surprisingly low
upper-limits ($< 100$ cm/s at depth $30$ Mm) for the large-scale
convection motions in the solar
convection zone. It is a possibility that solar convection
models over-estimate by a few orders of magnitude the deep convective
flow. For instance, an inadequate cut in the turbulent
scales due to finite numerical resolution alters the spectral
repartition of energy and hence may lead to incorrect estimates of the
large-scale convective flows. Another related possibility is that convection
may actually be driven by the strong cooling layer at the solar surface
\citep{Spruit:1997ul}, through the so-called \textit{entropy rain} of small
turbulent scales. On the other hand,
more recent observational results using ring-diagram analysis were
recently obtained by \citet{Greer:2015ff}, who found convective
flows two orders of magnitude faster than \citet{Hanasoge:2010jd} at
depth $30$ Mm. More cross-comparisons between helioseismology
techniques used to estimate convective flows in the Sun are today
needed to carefully pin down those observational constraints.

There is yet a somewhat more fundamental issue regarding our
understanding of stellar
convective turbulence, disregarding temporarily the exact mechanism exciting
it. Stellar interiors are mainly composed of stratified, fully ionized
hydrogen which can be modelled as a one-fluid plasma under the
magnetohydrodynamic (MHD) approximation (which is essentially composed of
a form of the Navier-Stokes equations combined with the heat transport
equation and coupled to an induction
equation). The microscopic dissipation capabilities of the stellar
plasma are relatively low: in the Sun, the microscopic viscosity in
the convection zone
typically lies in the $[1,10^2]$ cm$^2$/s range \citep[see,
\textit{e.g.},][]{Miesch:2005wz}. Considering the
lowest large-scale convective velocities estimates from
\citet{Hanasoge:2012fr} ($v \simeq [10^2,10^3]$ cm/s), and a
typical solar convection zone depth
$d\simeq 0.3\,R_\odot$, the typical Reynolds number of the large-scale
flows in the solar convection zone is at the very least 
$Re = v d / \nu \gtrsim 10^{10}$. Such a tremendously
high Reynolds number is, for the time being, unfortunately
inaccessible to both numerical simulations and laboratory
experiments. As a result, the spectral repartition of energy
(large-scales \textit{vs} small-scales, direct \textit{vs} inverse
cascades of energy) in the solar convection zone is today
unknown. The microscopic Prandtl number in the solar convection zone
($Pr=\nu/\kappa \lesssim 10^{-3}$) is
furthermore very challenging to reach with current numerical simulations,
since it implies a difference of at least three orders of magnitude for
viscous and heat dissipation time-scales. 
Finally, stellar convection zones are also
magnetized and are thought to generally support dynamo action, making
their theoretical and numerical modelling an outstanding challenge. 

In order to approach the extreme parameters of solar (and more
generally stellar) convection, \reff{which are unreachable with
  present computational power,}
various numerical simulation techniques have been developed. Two
complementary paths can generally be followed.
\reff{First, numerical simulations
can be designed on localized, small portions of the solar convection
zone \citep[\textit{e.g.}][and references
therein]{Rempel:2014db,Kitiashvili:2015hp}.} Even with this approach,
solar parameters are extremely hard to
achieve. Furthermore, the problem is truncated at the largest scales
(due to the small box extent) and hence cannot address the important
issue of large-scale convective motions and the sustainment of 
differential rotation or large-scale magnetism. A second path may
thus be followed, where the global convection zone is modelled but the
small-scales are parametrized. \reff{Both paths follow} the general approach of
large-eddy simulations (LES), on which we will focus in this work.

The simplest parametrization of sub-grid scales (SGS) consists in modelling
their effects as enhanced viscosity and heat diffusivity. This
approach has the advantage of giving the modeller full control on the sub-grid
scales model, but lacks a physical justification in the context
of solar (and stellar) convection. In contrast to classical
turbulence, an inertial range for convective
turbulence can be defined as the range of scales where the non-linear, local
advective energy transport balances the buoyancy source of
convection or the turbulent pressure gradient \citep[see,
\textit{e.g.},][]{Bolgiano:1959fv,Rincon:2006jm}. If the
smallest resolved scales of
the LES model lie within this modified inertial range of the turbulent
spectrum, the so-called dynamic Smagorinsky
procedure \citep[see][]{Smagorinsky:1963jb,Germano:1991cg} can
be used to mimic a given
self-similar spectrum for the smallest resolved scales of the model
(see \citealp {Nelson:2013fa}  for an implementation of such a method
in the context of solar
convection). Finally, another approach has been
pursued with the so-called implicit-LES (ILES) methods. Indeed, no
physically-rooted SGS model of the full MHD equations exists today
\citep[for recent advances in this direction, see][]{Yokoi:2013di,Chernyshov:2014kt}. A
pragmatic approach can hence consist in minimizing (\textit{e.g.}
down to the numerical stability limit) the effect of the
unresolved scales on the scales resolved by the model. The advantage
of this method is that it ensures that the effect of sub-grid scales
is minimized for a given grid size. However, the SGS model is fully
subsumed by the numerical method, and its role in the development of
the large scales is, unlike for explicit LES, non-trivial to estimate \textit{a posteriori}.

The aim of this work is to compare comprehensively the ILES and LES
modelling techniques for an idealized turbulent convection zone. We
design a turbulent convection zone simulation based on the anelastic
benchmark of \citet{Jones:2011in}. We use the EULAG code
\citep{Prusa:2008df} to compute ILES based on
the MPDATA algorithm \citep[see,
\textit{e.g.}][]{Smolarkiewicz:2013hq}, and use the ASH code
\citep{Clune:1999vd,Miesch:2000gs,Brun:2004ji} to compute the LES
counterparts. In a
pioneering work, \citet{Elliott:2002ez} first attempted
to reconcile empirically simulations of convective turbulence carried
with LES and ILES. Here, we quantify the dissipation of the ILES with
an original analysis of the energy transfers in spectral space. We
show that the effect of the SGS modelling
in the ILES can be interpreted in terms of enhanced viscosity and
diffusivity operators similar to LES.

The paper is organized as follows. The set of anelastic equations used
to describe convective turbulence in stars is given in Section
\ref{sec:modell-stell-deep}, along with a description of the two
numerical codes (EULAG and ASH) used in this work. A fiducial
convective turbulence
simulation in spherical geometry with the EULAG code is presented in
Section \ref{sec:fiduc-stell-conv}. In Section
\ref{sec:estimation-sub-grid}, we develop an original method
based on energy transfers in the spectral spherical
harmonics space to estimate effective dissipation coefficients. These
coefficients are then used in an ASH LES that
satisfyingly mimics the EULAG ILES (Section
\ref{sec:validation-with-ash}). We finally conclude and summarize our
results in Section \,\ref{sec:conclusions}.

\section{Modelling stellar deep convection}
\label{sec:modell-stell-deep}

\subsection{Formulation of the anelastic equations}
\label{sec:form-anel-equat}

The equations solved by ASH and EULAG are the Lantz-Braginsky-Roberts
\citep[LBR, see][]{Lantz:1999hm,Braginsky:1995kd} ---or,
equivalently, the Lipps-Hemler \citep{Lipps:1982km,Lipps:1985kg}--- set of anelastic equations
(see \citealp{Vasil:2013ij} for a recent review on sets of anelastic  equations).
The perturbed equations are written with respect to an ambient state
(hereafter denoted with the subscript $a$) that theoretically may
differ from the background state (hereafter denoted with bars) around
which the generic anelastic equations
were derived \reff{(both states will be detailed in Section \ref{sec:backgr-ambi-stat})}. The
background state is supposed to be isentropic, and both the ambient and
background states are supposed to satisfy the hydrostatic equilibrium.
The anelastic equations written in the stellar rotating frame $\boldsymbol{\Omega}_\star$ are
\begin{align}
 \label{eq:LBR}
 \nabla\cdot\left(\bar{\rho}\mathbf{u}\right) & =  0 \, , \\
 \label{eq:LBRE_u}\mbox{D}_{t} \mathbf{u} &= -\nabla
 \left(\frac{p}{\bar{\rho}}\right) - \frac{S}{c_{p}}\mathbf{g} 
 - 2\boldsymbol{\Omega}_\star\times\mathbf{u} + \frac{1}{\bar{\rho}}\boldsymbol{\nabla}\cdot \boldsymbol{\mathcal{D}}\, , \\
\label{eq:LBRE_S}\mbox{D}_{t} S &= 
-\left(\mathbf{u}\cdot\nabla\right)S_{a} -\frac{S}{\tau} + 
Q_{\kappa} \, , 
\end{align}
where the perturbed quantities are denoted without prime for the sake
of simplicity, and $\mbox{D}_{t}$ is the
material derivative. We recall that we use standard notation for the
basic fluid quantities, \textit{i.e.} ${\bf u}$ is the fluid velocity,
$\rho$ its density, $p$ its pressure, and $S$ its specific entropy. The dissipative terms, when present, are defined by 
\begin{align}
  \label{eq:visous_stress_tensor}
  \boldsymbol{\mathcal{D}} =& -2\bar{\rho}\nu\left(\boldsymbol{\epsilon} -
    \boldsymbol{I}\frac{\nabla\cdot\mathbf{u}}{3}\right)\, , \\
  \label{eq:heat_dissipation}
  Q_{\kappa} =& \frac{1}{\bar{\rho}\bar{T}}\nabla\cdot\left(\kappa 
\bar{\rho}\bar{T}\nabla S\right) \, ,
\end{align}
where $\boldsymbol{\epsilon} = \left(\nabla{\bf u} +
  \left(\nabla{\bf u}\right)^T\right)/2$ is the strain
rate tensor and $\boldsymbol{I}$ the identity tensor. 
Note that here we choose not to consider viscous heating, \refff{as in
ILES with EULAG there is no equivalent in the implicit treatment of
sub-grid scales.} 
\reff{This choice has the potential drawback of not
formally conserving the total energy in the system.} 
In addition, we use a standard perfect gas equation of state which is
linearized around the background state. 

In the preceding equations,
convection is forced by the conjunction of the advection of the
unstable ambient entropy profile $S_a$, and a Newtonian cooling term
with a
characteristic timescale $\tau$ \citep[for
details,
see][]{Prusa:2008df,Smolarkiewicz:2013hq}. The Newtonian cooling damps
entropy perturbations over the timescale
$\tau$ which is always chosen to exceed the convective overturning
time. This ensures that on long time-scales, the model mimics a stellar
convection zone remaining in thermal equilibrium
\citep[\textit{e.g.}][]{Cossette:2016vc}. \reff{We have modified the
  equations solved in ASH to include this Newtonian cooling term in
  order to force convection in exactly the same way in both codes. As
  a result, the same set of anelastic equations are solved in both
  cases, with the exception of explicit dissipation operators in ASH.}

The anelastic equations can equivalently be specified in terms of
potential temperature $\Theta$, which is related to the specific entropy through
\begin{align}
  \label{eq:entropy_pot_T}
  c_{p}\ln\bar{\Theta} =& \bar{S}\, , \\
  \label{eq:entropy_pot_T2}
  \frac{\Theta}{\bar{\Theta}} =& \frac{S}{c_{p}}\, .
\end{align}
Equations (\ref{eq:LBRE_u}) and (\ref{eq:LBRE_S}) can be written in terms of potential temperature through
\begin{align}
  \label{eq:LBRET_u}\mbox{D}_{t} \mathbf{u} =& -\nabla
  \left(\frac{p}{\bar{\rho}}\right) - \frac{\Theta}{\bar{\Theta}}\mathbf{g} -
  2\boldsymbol{\Omega}_\star\times\mathbf{u} +
  \frac{1}{\bar{\rho}}\boldsymbol{\nabla}\cdot \boldsymbol{\mathcal{D}}\, , \\
  \label{eq:LBRET_T}\mbox{D}_{t} \Theta =& 
-\left(\mathbf{u}\cdot\nabla\right)\Theta_{a} - \frac{\Theta}{\tau}
+ \frac{1}{\bar{\rho}\bar{T}}\nabla\cdot\left(\kappa
\bar{\rho}\bar{T}\nabla \Theta\right) \, .
\end{align}
Note that the ambient temperature $\Theta_a$ is formally defined similarly to
$\bar{\Theta}$, \textit{i.e.} $c_p\ln\Theta_a \equiv S_a$. When deriving
Equation (\ref{eq:LBRET_T}) from (\ref{eq:LBRE_S}) we have assumed
that $\left({\bf u}\cdot\nabla\right)S_a \simeq 
  \left({\bf u}\cdot\nabla\right) c_p\Theta_a/\bar{\Theta}$, which is a reasonable
assumption given that the ambient entropy profile differs only by a small amount from the background entropy profile (see hereafter in
Section \ref{sec:backgr-ambi-stat}).

\subsection{Background and ambient states}
\label{sec:backgr-ambi-stat}

Our numerical setup closely follows the anelastic benchmark of
\citet{Jones:2011in}. We consider a spherical shell of aspect ratio $\beta=R_{i}/R_{\star}$, and
we note $d = R_{\star}-R_{i} = R_{\star}(1-\beta)$. We assume a gravity profile $g=GM/r^{2}$, for which the
anelastic equations admit an equilibrium (denoted with bars)
polytropic solution \citep[see, \textit{e.g.},][]{Jones:2011in}
\begin{align}
  \label{eq:polytrop}
  \bar{\rho} = \rho_{c}\xi^{n}, \,\, \bar{P} &=
  P_{c}\xi^{n+1},\,\,  \bar{T} = T_{c}\xi\, , \\ \xi &= c_{0} +
                                                       \frac{c_{1}d}{r}\, ,
  \label{eq:polytrop2}
\end{align}
where $n$ is the polytropic index, $\rho_c, P_c, T_c$ are the
density, pressure and temperature at the bottom of the domain, and the constants $c_{0}$ and
$c_{1}$ are given by
\begin{align}
  \label{eq:csts}
  c_{o} =& \frac{2\alpha-\beta-1}{1-\beta} , \, \, c_{1} =
  \frac{(1+\beta)(1-\alpha)}{(1-\beta)^{2}},\\
  \alpha =& \frac{\beta+1}{\beta \exp\left(N_{\rho}/n\right)+1} \, ,
\end{align}
where $N_{\rho}=\ln(\rho_{i}/\rho_{o})$ is the number of density scale
heights in the layer. 
The background entropy
profile is given by
\begin{align}
  \label{eq:entropy}
  \bar{S} =
  c_{p}\ln\left(\frac{\bar{P}^{1/\gamma}}{\bar{\rho}}\right) = c_{p}\ln\left(\frac{P_c^{1/\gamma}}{\rho_c}\xi^{(n+1-n\gamma)/\gamma}\right) \, ,
\end{align}
where the standard adiabatic
exponent for a perfect gas is $\gamma=c_{p}/c_{v}=5/3$.
We choose a polytropic exponent $n=3/2$ to naturally ensure an
isentropic background state.

The ambient state needs to be specified only in terms of entropy and
potential temperature. The entropy jump throughout the domain, $\Delta
S$, is used to define the ambient entropy profile by
\begin{align}
  S_{a}(r) = \bar{S} + \Delta S \frac{\xi^{-n}(R_{\star}) -
    \xi^{-n}(r)}{\xi^{-n}(R_{\star}) - \xi^{-n}(R_{i})}\, ,
\end{align}
which we recall is related to the aforementioned ambient potential temperature profile by
$\Theta_a = \exp{\left(S_a/c_p\right)}$.
 
It should be noted that the background and ambient states used in this work
differ from the ones used in past published ASH and EULAG
simulations. We implemented those profiles in EULAG to be able to
compare easily our results with ASH simulations. Furthermore, in this
work only the convective layer is modelled, with no underlying stable
layer.

\subsection{Numerical methods}
\label{sec:numerical-methods}

We use two codes based on different numerical methods.

The Eulerian-Lagrangian (EULAG) code is designed to use either Eulerian (flux form) or semi-Lagrangian
(advective form) integration schemes \citep[see][]{Prusa:2008df,Smolarkiewicz:2013hq}.
In the case presented here, Equations (\ref{eq:entropy_pot_T}) and (\ref{eq:LBRET_u})
are written as a set of Eulerian conservation laws and projected on a
geospherical coordinate system \citep{Prusa:2003fa}. 
EULAG solves the evolution equations
using MPDATA (multidimensional positive definite advection transport 
algorithm), which belongs to the class of nonoscillatory Lax-Wendroff 
schemes \citep{Smolarkiewicz:2006dz}, and is more specifically a second-order-accurate
nonoscillatory forward-in-time template. Since all dissipation is delegated to MPDATA,
this provides an implicit turbulence model
\citep{Domaradzki:2003cb}. Implicit dissipation diminishes if
explicit dissipation is introduced in the model, providing seamless
transition between ILES and LES \citep[see][and references
therein]{Margolin:2006dg}. In EULAG, all linear forcing terms are
integrated in time using a second-order Crank-Nicholson scheme.

The Anelastic Spherical Harmonics (ASH) code is a pseudo-spectral code
\citep[see][]{Boyd:1989wz,Glatzmaier:1984jh,Clune:1999vd} based
on a spherical harmonics decomposition, which avoids the classical
issues related to the convergence of meridians at the poles of a
sphere. As in EULAG, the linear terms of the anelastic equations are
treated with an
implicit Crank-Nicholson scheme of order 2. An Adams-Bashford
scheme is used for the non-linear terms. The latter are evaluated in
physical space, making the numerical method overall 
\textit{pseudo-spectral}. In the radial direction, variables can be
described either \textit{via} a Chebyshev decomposition, or
\textit{via} a finite difference method \citep{Alvan:2014gx}.
We use the latter here with a fourth-order finite difference scheme.

\begin{table}[tb]
  \centering
  \begin{tabular}{lc}
    \hline
    Parameter &  Value\\
    \hline
    $R_\star$ [$R_\odot$] & 1 \\
    $\beta = R_i/R_\star$ & 0.7 \\
    $M$ [$M_\odot$] & 1 \\
    $\Omega _\star$ [10$^{-6}$ rad s$^{-1}$] & 6.488 \\
    $G$ [dyne-cm$^{2}$ g$^{-2}$] & 6.673 $\times$ 10$^{-8}$ \\
    $c_P$ [erg g$^{-1}$ K$^{-1}$] & 3.4 $\times$ 10$^8$ \\
    $\rho_i$ [g cm$^{-3}$] & 0.2 \\
    $N_\rho = \ln\left(\rho_i/\rho_o\right)$ & 1.5\\
    Polytropic exponent $n$ & 3/2 \\
    $\tau$ [s] & 5.184 $\times$ 10$^7$ \\
    $\Delta \, S$ [erg g$^{-1}$ K$^{-1}$] & 2 $\times$ $10^{3}$\\
    \hline
  \end{tabular}
  \caption{Fiducial stellar convection zone parameters. The solar
    radius is $R_\odot = 6.9599 \times 10^{10}$ cm, and the solar mass
    $M_\odot = 1.99 \times 10^{33}$ g.}
  \label{tab:params_case}
\end{table}

\reff{We consider in both codes
stress-free, impermeable boundaries at the top and bottom of
the domain such that}
\begin{align}
  \label{eq:v_boundaries}
  u_{r} = \partial_{r} \left(u_{\theta}/r\right) = \partial_{r} \left(
    u_{\varphi}/r\right) = 0\, \mbox{at } r=R_{i}, R_{\star}\, .
\end{align}
The entropy gradient is set to zero on the upper and lower
boundaries. The simulations presented in this work are initialized with
random, small-amplitude perturbations around the background profiles
defined in Section \ref{sec:backgr-ambi-stat}.
 
\section{Fiducial stellar convection zone with EULAG}
\label{sec:fiduc-stell-conv}

\begin{figure}[!htb]
  \centering
  \includegraphics[width=\linewidth]{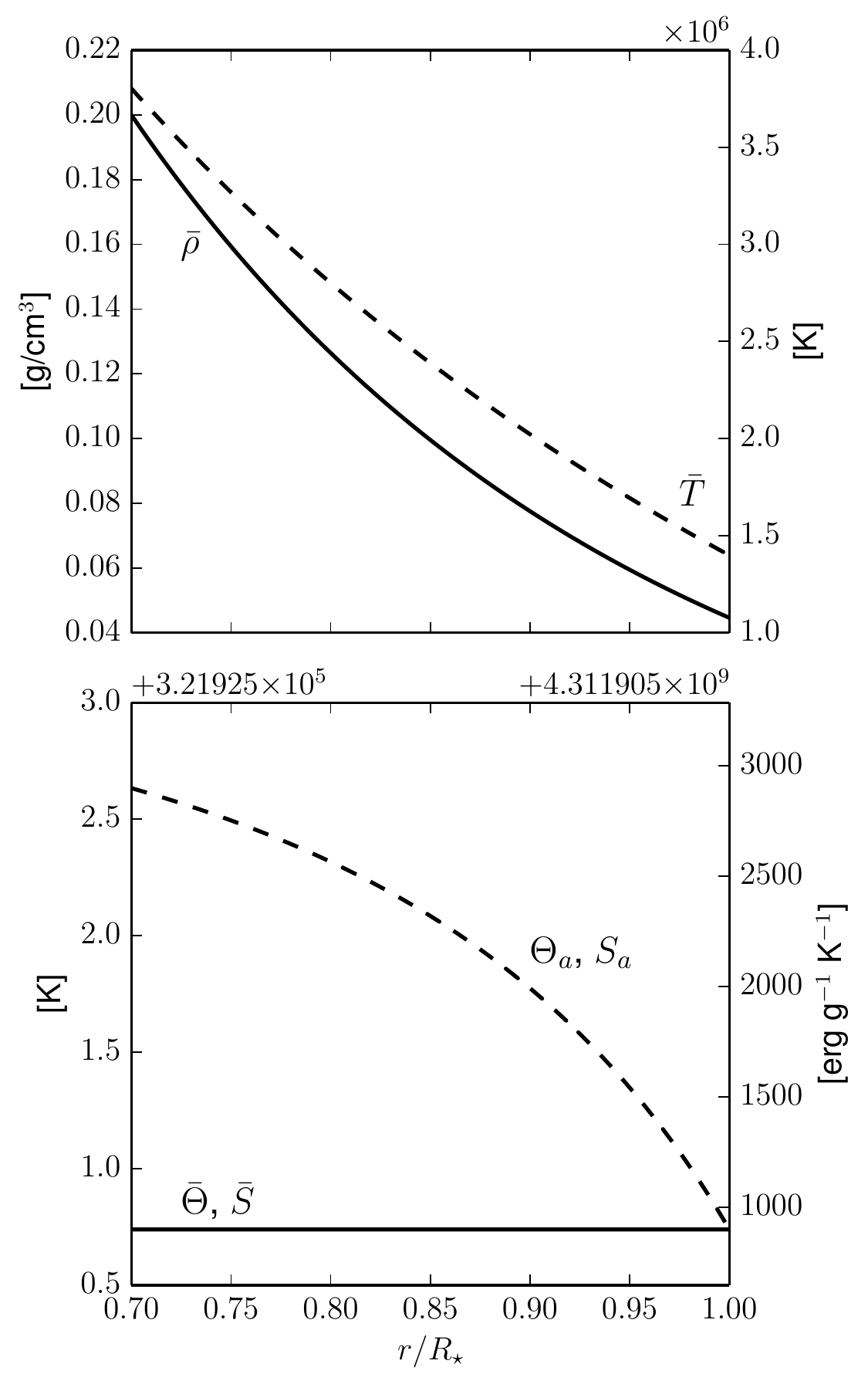}
  \caption{\textit{Top panel}: Background density $\bar{\rho}$ (left
    vertical axis) and
    temperature $\bar{T}$ (right vertical axis) profiles as a function
    of spherical
    radius. \textit{Bottom panel}: Background and ambient potential
    temperature and entropy profiles. The left vertical axis
    corresponds to the potential temperature, and the right axis
    to the entropy.}
  \label{profiles}
\end{figure}
 We consider a fiducial stellar convection zone computed with the EULAG
code. The parameters of our fiducial case are indicated in Table
\ref{tab:params_case}. We consider a
convection zone with a solar-like aspect ratio and which rotates $2.4$
times faster than the Sun, which ensures that the Rossby number \reff{$R_o =
\omega/2\Omega_0$ (where $\omega$ is the average vorticity in the
middle of the convection zone)} remains
sufficiently smaller than one, and consequently that the model is strongly
influenced by rotation. The
background density and temperature profiles are shown in the upper
panel of Figure \ref{profiles}. A moderate density contrast (1.5
density scale-heights) is adopted to limit the size of the smallest
turbulent scales near the top of the domain. In the lower panel we show the
background and ambient potential temperature (and equivalently
entropy) profiles. A moderate entropy constrast over the convective
shell is chosen to ensure the model is above the critical onset of
convection, but remains in an accessible turbulent regime.

The numerical method in EULAG implicitly supplies
the dissipation needed to maintain numerical stability, which in turns
varies with the grid resolution. We
consider two physically identical cases, labelled E1 and E2, in which the grid
resolution is respectively $N_r \times N_\theta \times N_\varphi$ = 51
$\times$ 64 $\times$ 128, and 101
$\times$ 128 $\times$ 256, with respective time steps of $900$ and
$450$ seconds. By increasing the resolution (in both time and space) by a factor of 2, the implicit
dissipation of the numerical scheme of EULAG is expected to be reduced
by a factor of 4 \citep[see, \textit{e.g.},][]{Prusa:2008df}. 
\begin{figure*}[!htb]
  \centering
  \includegraphics[width=0.64\linewidth]{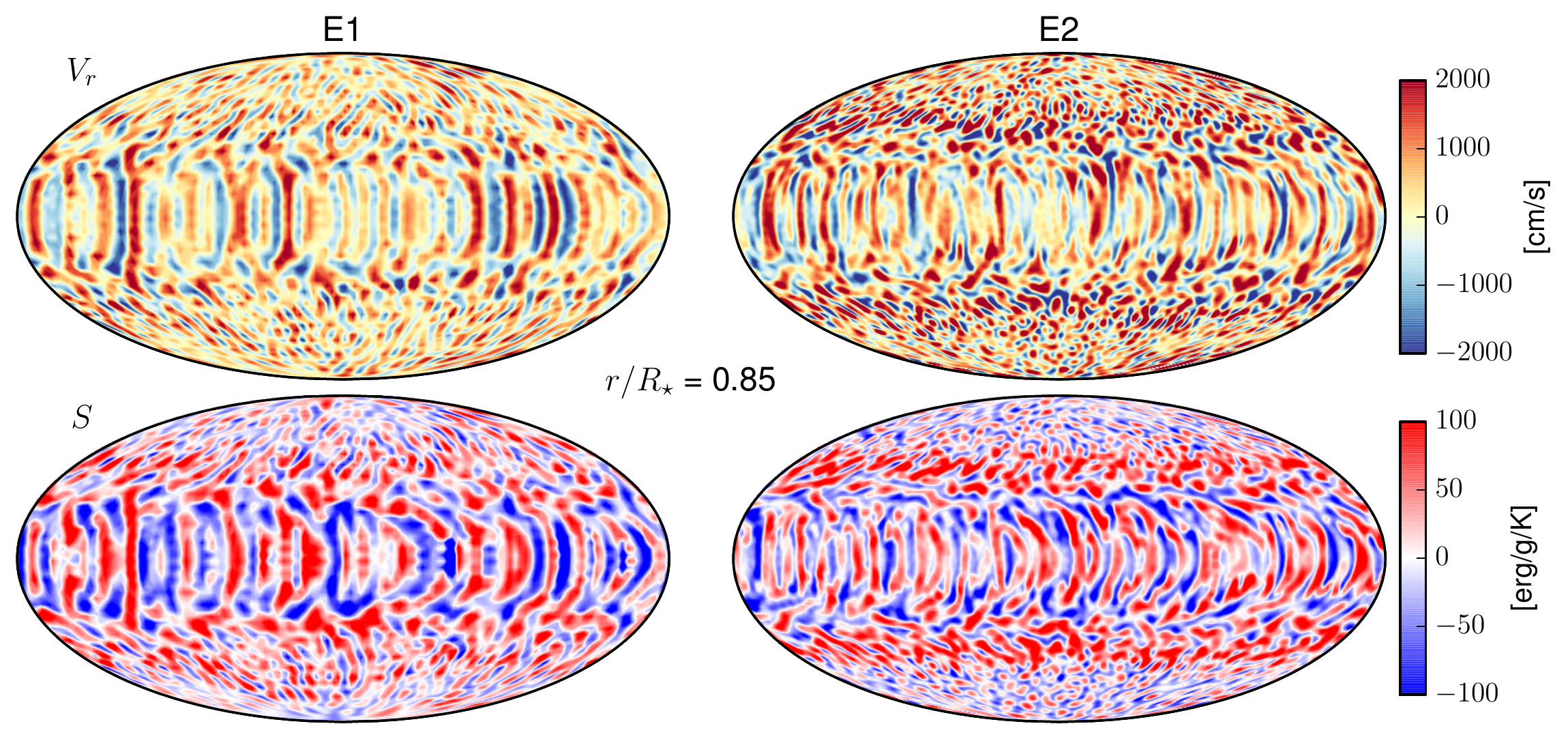}
  \includegraphics[width=0.35\linewidth]{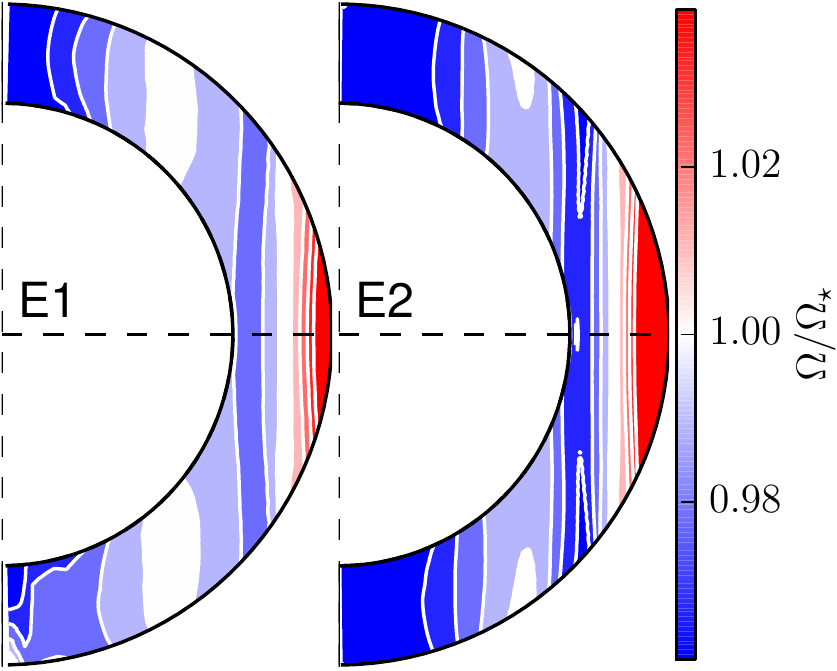}
  \caption{\textit{Left panels}: Mollweide projections in the middle of
    the convection zone of the radial velocity (upper panels) and
    entropy perturbations (lower panels) for cases E1 (left) and E2
    (right). The positive $v_r$ and $S$ perturbations are denoted in
    red, and the negative values in blue. \textit{Right panels}:
    Differential rotation ($\Omega$) profile (averaged over time and longitude)
    on the meridional plane for cases E1 (left) and E2 (right). The differential rotation profile is
    normalized to the stellar rotation rate $\Omega_\star$. Rotation
    faster than $\Omega_\star$ is shown in red, slower in blue, and
    co-rotation appears in white.}
  \label{fig:DR_Vr_S_Eulag}
\end{figure*}
 We display in Figure
\ref{fig:DR_Vr_S_Eulag} the radial velocity and entropy perturbations (left panels) in both cases
on a Mollweide projection in the middle of the convection zone. As
expected, smaller structures are observed when the resolution is increased. In both cases, the
so-called \textit{banana cells} clearly appear on the
equator. Interestingly, the convective luminosity almost does not change between
the two cases and peaks at about $0.12\, L_\odot$\footnote{The solar
  luminosity is $L_\odot = 3.846 \times 10^{33}$ erg/s} in the middle of
the convection zone. The convective kinetic energy only increases
by $10\%$ from E1 to E2, and the temperature perturbations
compensate this variation which leads to a very similar convective
luminosity. The main difference between the two cases hence lies in
the kinetic energy of the differential rotation, which results from
the complex interplay of the turbulent scales (through Reynolds
stresses) involved in the
simulation. The differential rotation (right panels) 
exhibits a solar-like (fast equator, slow
poles) pattern in both cases. Case E2 possesses a significantly
stronger pole-equator constrast ($\sim 6\%$) compared to E1 ($\sim 2\%$).
The iso-$\Omega$ contours are in both cases mostly parallel to the
rotation axis, and an almost co-rotating stripe is observed at mid-to
high latitude, indicating that the simulation is indeed in a regime strongly
influenced by rotation (low Rossby number).

\section{Estimation of the sub-grid scale modelling effects in EULAG}
\label{sec:estimation-sub-grid}

We now quantitatively estimate the effect of the numerical treatment
of sub-grid scales in EULAG. We base our analysis on the estimation of
energy transfers in the turbulent convection zone simulations. We
derive the spectral analysis method in Section
\,\ref{sec:spectral-analysis} and detail the results in Section \,\ref{sec:spectr-analys-result}.

\begin{figure*}[hbt]
  \centering
  \includegraphics[width=0.49\linewidth,page=1]{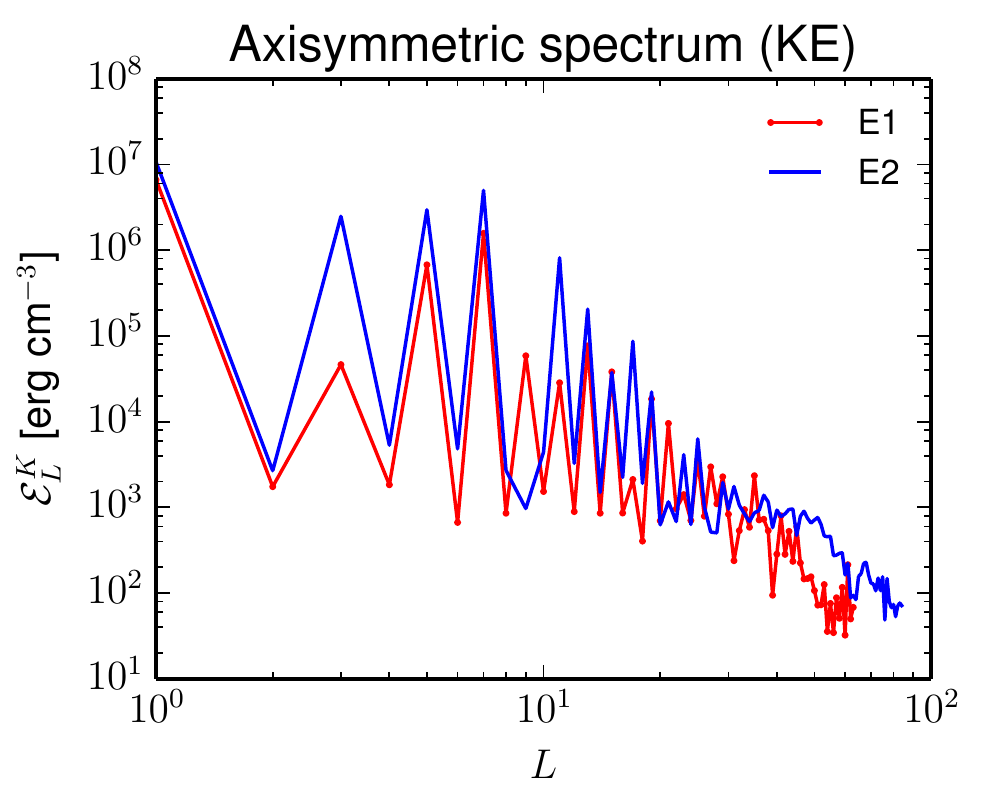} 
  \includegraphics[width=0.49\linewidth,page=1]{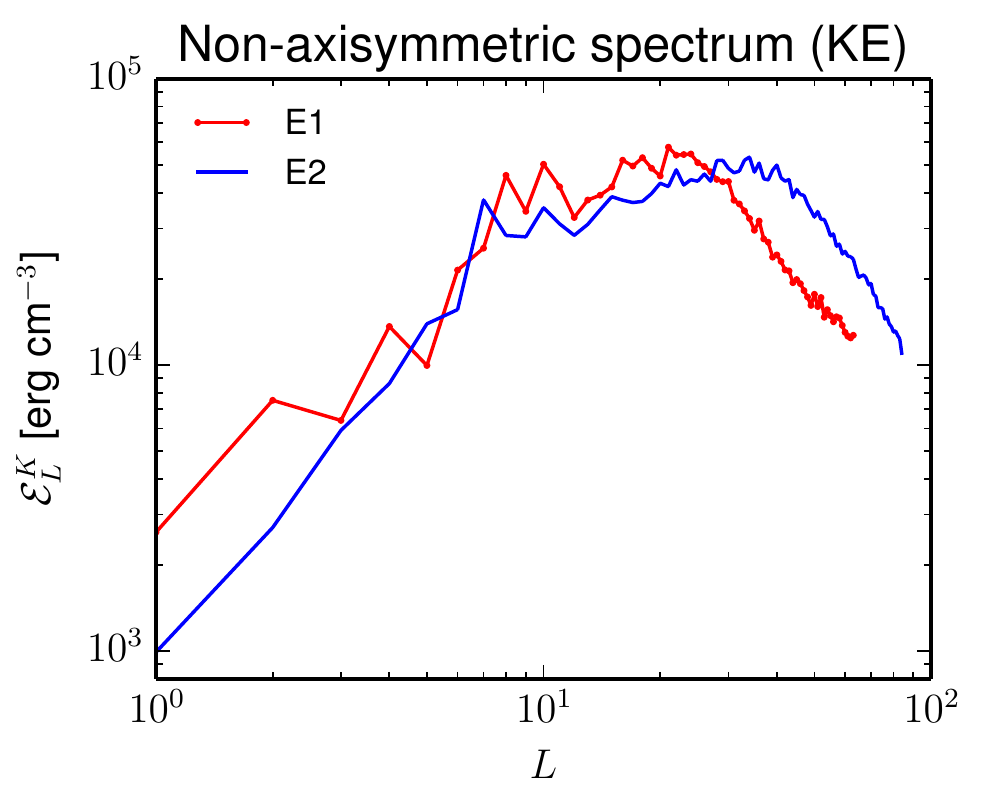} 
  \includegraphics[width=0.49\linewidth,page=2]{Asp_compare} 
  \includegraphics[width=0.49\linewidth,page=2]{Nsp_compare} 
  \caption{Axisymmetric (left panels) and non-axisymmetric (right
    panels) kinetic energy (upper panels) and entropy (lower panels)
    spectra averaged over 40 rotation periods. Case E1 spectra are
    shown in red and case E2 in blue. The
    spectra are summed over the spherical harmonics degree $m$ and
    displayed as a function of $L$ (see text).}
  \label{fig:spectra}
\end{figure*}

\subsection{Spectral Analysis: Method}
\label{sec:spectral-analysis}

We analyze the results from the EULAG code 
with the help of a spectral analysis of energy transfers
between scales in the spherical harmonics space. This method was
originally developed in \citet{Strugarek:2013kt} in the context of
characterizing the transfer of magnetic energy between scales in
dynamo simulations. It was
more recently adapted to the EULAG code to study the self-consistent development of
MHD instabilities in the solar tachocline \citep{Lawson:2015fq}. We
extend it here to study kinetic energy and entropy spectral transfers 
\citep[we refer the reader to][and to \ref{sec:spectr-transf-equat} for more details about this
spectral analysis method]{Strugarek:2013kt}. The scalar and vectorial
quantities are respectively projected on the standard and vectorial
spherical harmonics bases \citep{Rieutord:1987go}
\begin{align}
  \label{eq:SH_scal}
  Y_l^m = \sqrt{\frac{(2l+1)}{4\pi}\frac{(l-|m|)!}{(l+|m|)!}}P^{|m|}_l(\cos \theta)e^{im\varphi}  
\end{align}
and
\begin{align}
  \label{eq:RST}
  \left\{
  \begin{array}{lcl}
    \mathbf{R}_l^m &=& Y_l^m {\bf e}_r \\
    \mathbf{S}_l^m  &=& 
                         \partial_\theta Y_l^m {\bf e}_\theta +
                         \frac{1}{\sin{\theta}}\partial_\varphi Y_l^m
                         {\bf e}_\varphi\\
    \mathbf{T}_l^m  &=& 
                         \frac{1}{\sin{\theta}}\partial_\varphi
                         Y_l^m {\bf e}_\theta  -\partial_\theta Y_l^m
                         {\bf e}_\varphi
  \end{array}
  \right.\, ,
\end{align}
where $P^m_l$ are the Legendre polynomials. 

\begin{figure*}[htb]
  \centering
     \includegraphics[width=\linewidth]{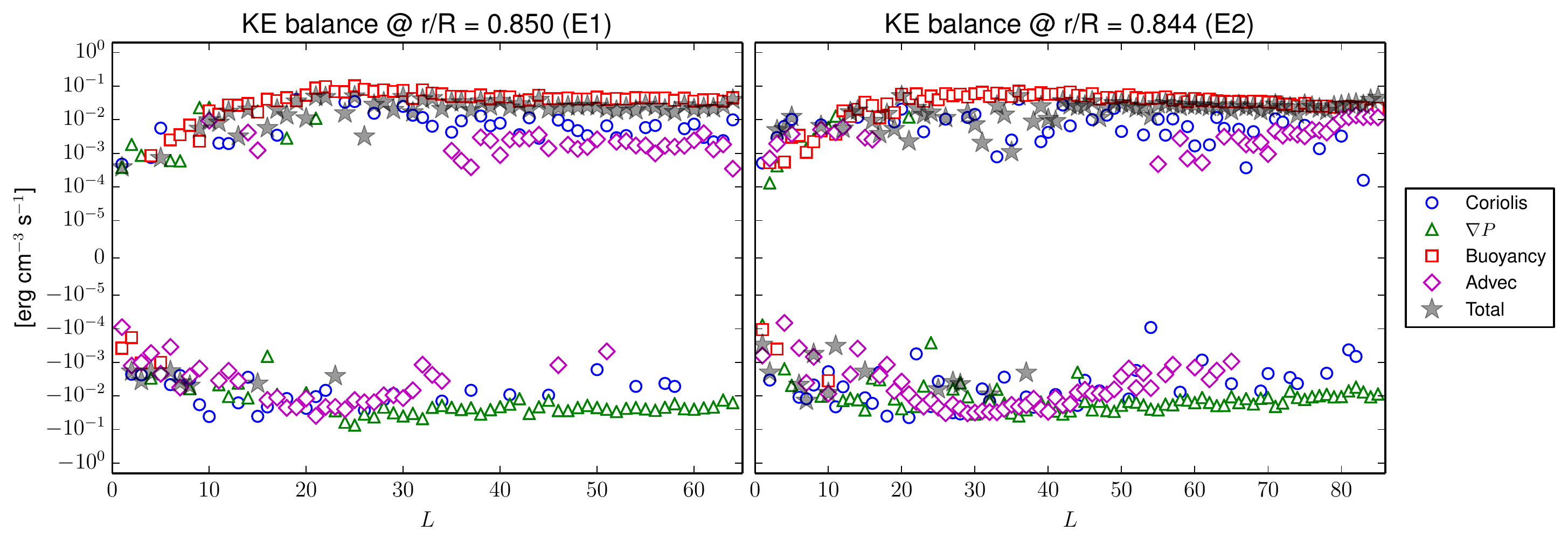} 
     \includegraphics[width=\linewidth]{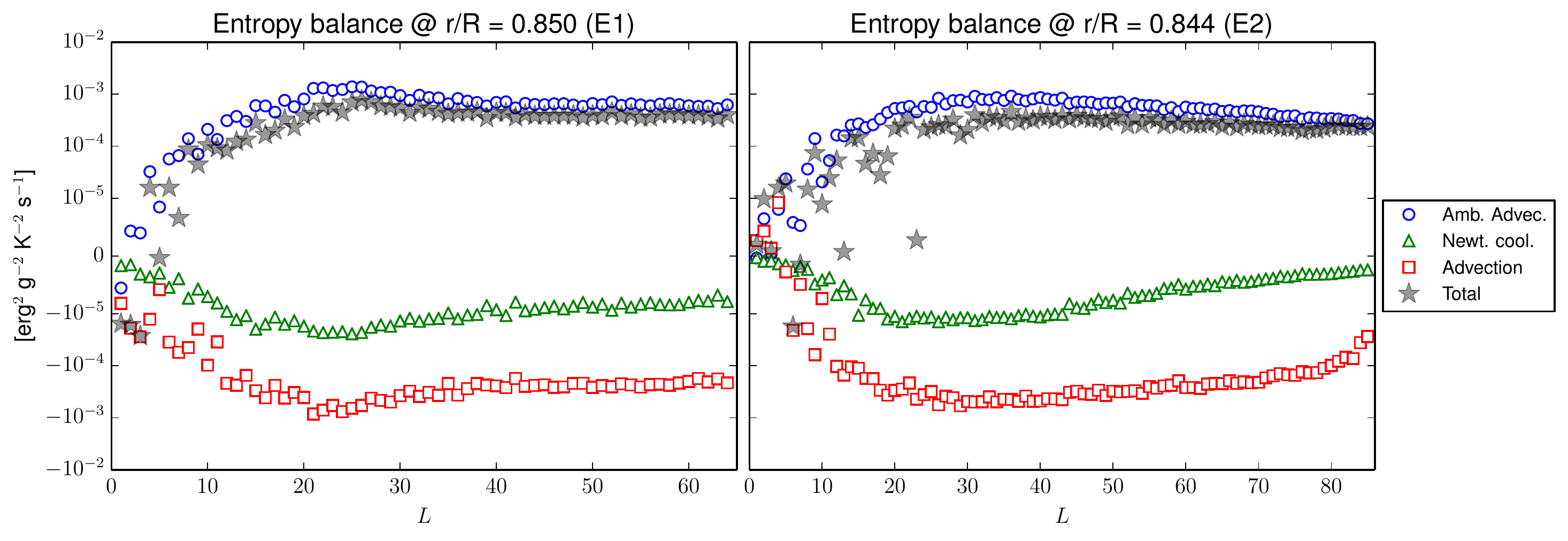}
  \caption{Kinetic energy (upper panels) and entropy (lower panels)
    balance Equations (\ref{eq:evol_KE})-(\ref{eq:evol_S}) for cases E1
    (left panels) and E2 (right panels), in the middle of the
    convection zone. For each equation the various terms are labeled
    with a coloured symbol, and the total of all the contributions is
    shown by the gray transparent stars.}
  \label{fig:nu_kappa_middle_cz}
\end{figure*}

We define the kinetic energy and entropy spectra by
\begin{align}
  \label{eq:KE_sp}
  \mathcal{E}^K_L(r) =& \frac{1}{2}\bar{\rho}\iint {\bf u}_L \cdot
                         {\bf u}_L^{cc} {\rm d}\Omega \, ,\\
  \mathcal{E}^S_L(r) =& \frac{1}{2}\iint S_L \cdot
                         S_L^{cc} {\rm d}\Omega
                         \, ,
  \label{eq:S_sp}
\end{align}
where ${\rm d}\Omega = \sin\theta{\rm d}\theta {\rm d}\varphi$ and the
exponent $^{cc}$ denotes the complex conjugate.
The subscript $L$ corresponds to the sum over the subset of spherical harmonics
coefficients $(l,m)$ with fixed $l$ and $m$ restricted to a chosen ensemble
$\mathcal{M}$. In this work we will consider the two ensembles
$\mathcal{M}_0 = \{m=0\}$ and $\mathcal{M}_\star = \{m\in[-l,l],\,
m\ne 0\}$. We display those axisymmetric and non-axisymmetric spectra at mid-depth
for cases E1 and E2 in Figure \ref{fig:spectra}. The axisymmetric kinetic energy
spectrum (upper left panel) is dominated by the
differential rotation that populates the odd $L$ components of the
spectrum. The entropy axisymmetric spectrum (lower left panel) is
conversely dominated by even $L$ components, which correspond to the
mean pole-equator temperature contrast that establishes itself in the simulations.
The non-axisymmetric kinetic energy spectrum (upper right
panel) exhibits a peak around \reff{$L\sim 25$} in case E1 (\reff{$L\sim 35$} in
case E2) which corresponds to the dominant convective structure that
can be observed in the Mollweide projection in Figure
\ref{fig:DR_Vr_S_Eulag}. As expected, the non-axisymmetric kinetic
energy spectrum at scales $L>20$ is
shifted to smaller scales (higher $L$'s) when the resolution is
increased, but the overall shape of the \reff{non-axisymmetric kinetic
  energy} spectrum remains the same. The
non-axisymmetric entropy spectrum peaks at \reff{$L\sim 25$} in case E1 and
\reff{$L\sim 30$} in case E2. \reff{When the resolution
  is doubled, the entropy spectrum is not shifted to smaller scales but the
  negative slope extends down to the smallest scales resolved in the domain.}

Using the anelastic Equations
(\ref{eq:LBRE_u}) and (\ref{eq:LBRE_S}) with no explicit dissipation, we obtain the following evolution equations 
\begin{align}
  \label{eq:evol_KE}
  \dot{\mathcal{E}}^K_L(r) =& \mathcal{P}_L +
                               \mathcal{G}_L +
                               \mathcal{C}_L
                           + \sum_{L_1,L_2}
                              \mathcal{R}_L\left(L_1,L_2\right)\, , \\
  \label{eq:evol_S}
  \dot{\mathcal{E}}^S_L(r) =& \mathcal{S}^a_L + \mathcal{N}_L +
                              \sum_{L_1,L_2}\mathcal{A}_L\left(L_1,L_2\right)\, .
\end{align}

In the
kinetic energy Equation (\ref{eq:evol_KE}), the various terms in the
right hand side
correspond to contributions from the pressure gradient, buoyancy,
Coriolis force and non-linear advection of momentum. Note that the
Coriolis contribution $\mathcal{C}_L$ vanishes when summed over all
scales $L$ as it should, but is able to spectrally
redistribute energy among neighbour shells (see also
\citealt{Augier:2013dz} and \ref{sec:coriolis-term}). In the entropy Equation
(\ref{eq:evol_S}), they correspond to the ambient state advection,
Newtonian cooling, and non-linear advection of the entropy
perturbations. The detailed expressions
of the different terms can be found in
\ref{sec:kinetic-energy} and \ref{sec:entropy-equation}. We show in
Figure \ref{fig:nu_kappa_middle_cz} the various terms of Equations
(\ref{eq:evol_KE}) and (\ref{eq:evol_S}) for cases E1 and E2 at
mid-depth (we focus here solely on the non-axisymmetric spectra). The
different contributions are averaged over more than 40 stellar rotations
after the simulations have reached a steady-state (\textit{i.e.} after
the total kinetic energy is stabilized). In
the kinetic energy balance (upper panels), buoyancy (red squares)
is clearly the source of energy for almost all turbulent scales and is generally
opposed by the pressure gradient (green triangles). The non-linear advection
contribution (magenta diamonds) changes sign at the
peak of the spectrum because of a direct energy cascade: scales
larger than the maximum energy scale lose energy to the smaller
scales. The Coriolis force (blue circles) efficiently acts on the largest
scales due to the small latitudinal extent of the higher $L$
modes, as expected. 

The entropy balance is shown in the lower panels. The entropy perturbations
draw energy from the ambient entropy profile at all scales (blue circles) and are
stabilized by the non-linear advection (red squares). The Newtonian
cooling term (green triangles) contributes very marginally to the
entropy balance since we chose a long cooling timescale $\tau$ (see
Table \ref{tab:params_case}). We
finally note that no cascading process is observed in
the entropy transfers in these numerical experiments: the non-linear
advection is negative for all the resolved scales.

The total right hand side of the kinetic energy and entropy evolution
equations ---shown as gray stars in
Figure \ref{fig:nu_kappa_middle_cz}--- should cancel out in
steady-state. In Figure \ref{fig:nu_kappa_middle_cz}, we recall that the various
contributions were averaged over a time
period of 40 stellar rotations during which the total energy in the
system has stabilized. Hence, the system is in steady-state and the imbalance
(gray stars) can only be attributed to
the effect of the numerical scheme that we do not represent in these
plots and that tends to dissipate energy in
EULAG. We now propose a possible interpretation of this imbalance to obtain
quantitative estimates of the effect of the sub-grid scales modelling
in EULAG.

\subsection{Effective dissipation coefficients}
\label{sec:spectr-analys-result}

\begin{figure*}[htbp]
  \centering
     \includegraphics[width=\linewidth]{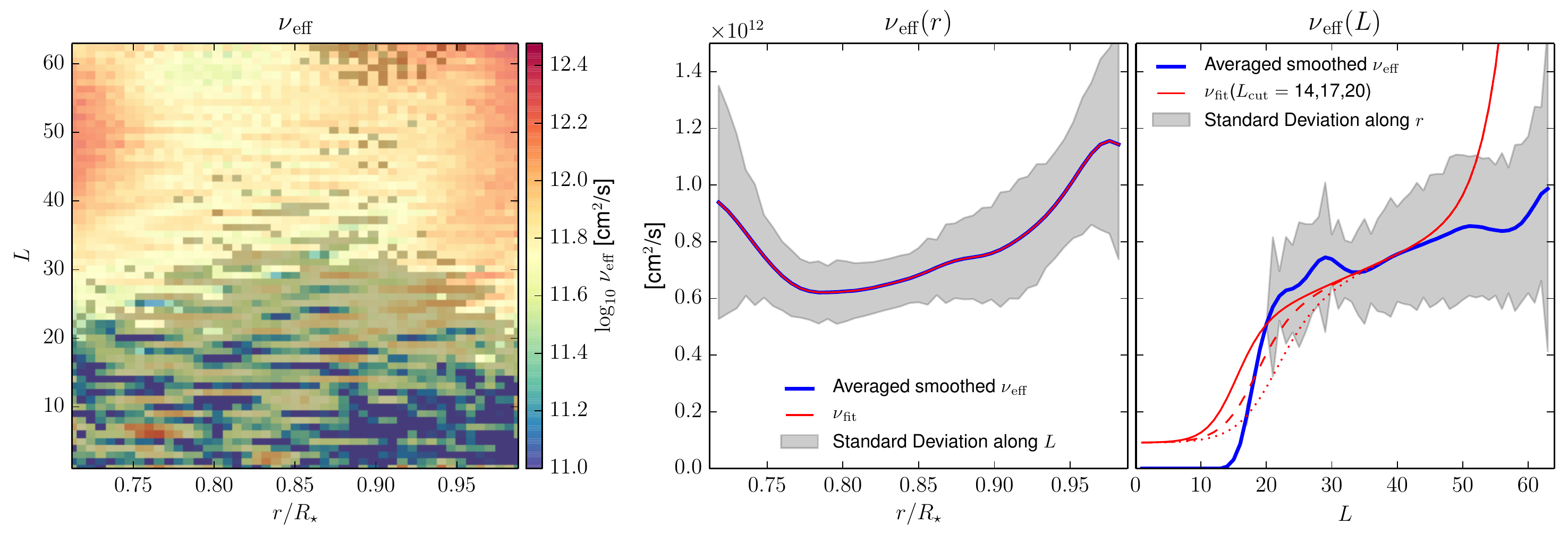} 
     \includegraphics[width=\linewidth]{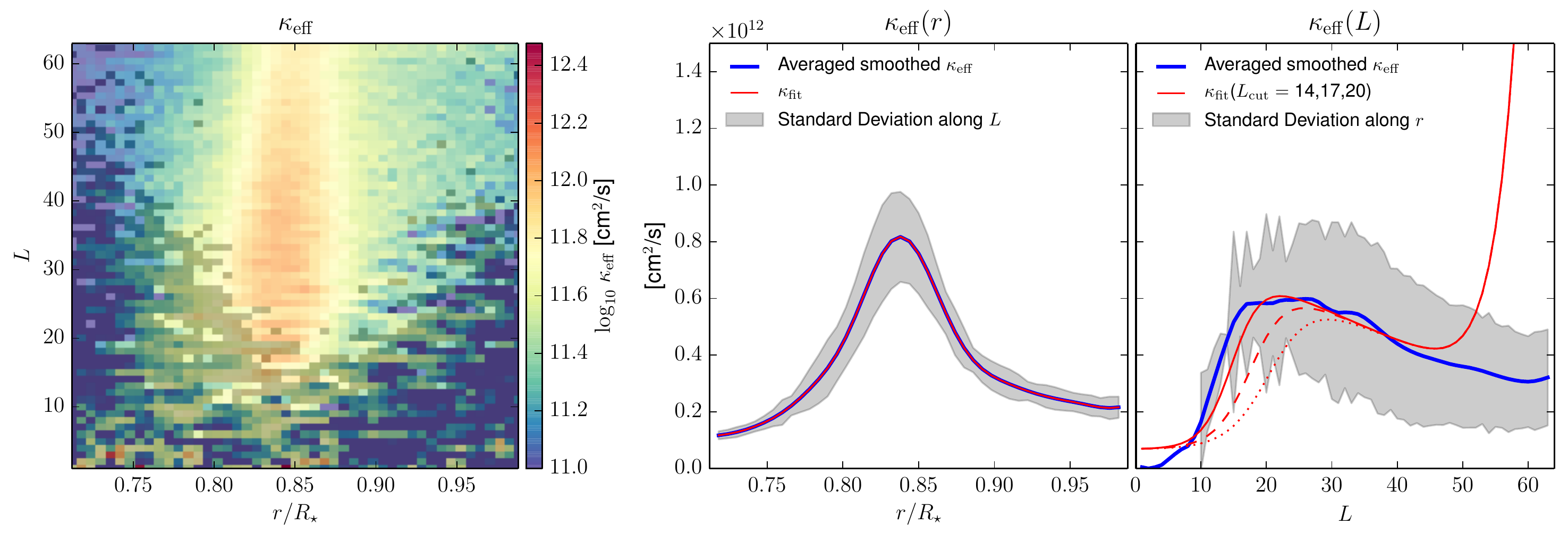}
     \includegraphics[width=\linewidth]{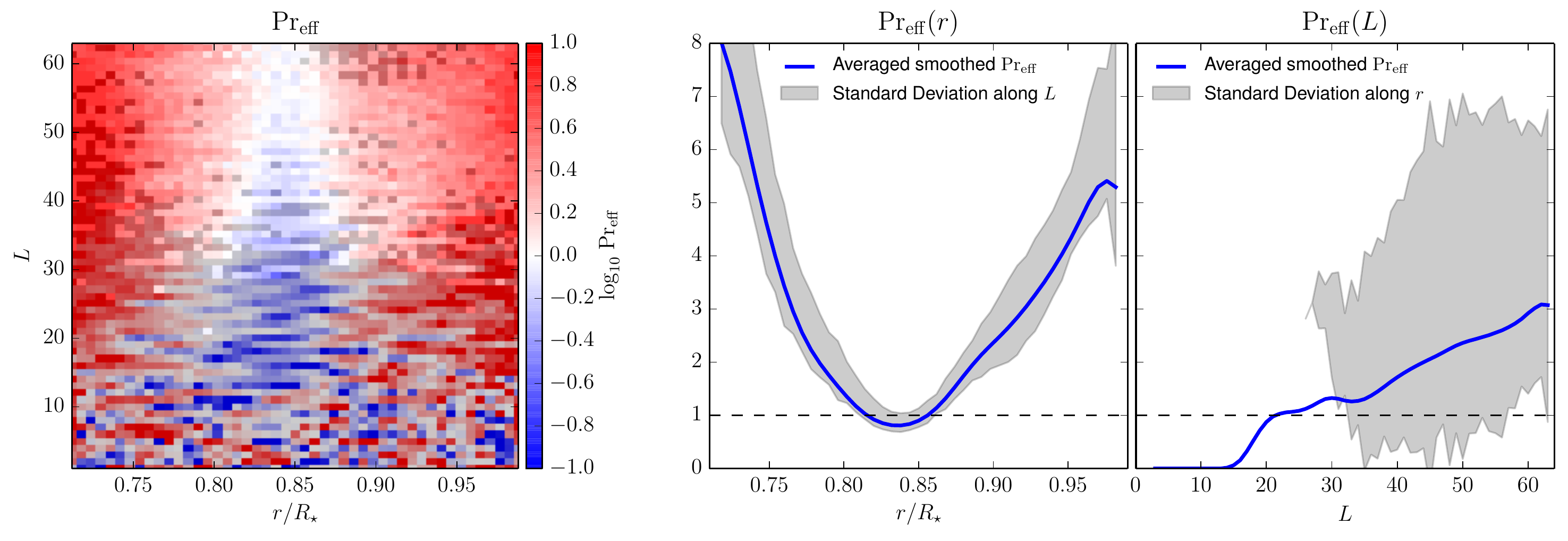}
  \caption{\textit{Left panels}: Effective visocity $\nu_{\rm eff}$
    (upper panel), heat dissipation coefficient $\kappa_{\rm eff}$
    (middle panel) and Prandtl number $Pr$ (lower panel), as a
    function of spherical radius and spherical harmonics degree
    $L$. The colormap in each panel is displayed on a logarithmic
    scale. In each panel, gray pixels label non-robust effective
    dissipation coefficients (see text). Note that in the lower panel,
    white areas correspond to an effective Prandtl number of unity. \textit{Right panels}:
    Averaged, smoothed profiles (blue lines) as a function of
    spherical radius (left) and spherical harmonics degree $L$
    (right). The gray area corresponds to the standard deviation of
    the effective dissipation coefficients around the average
    value. The red lines denote the fitted dissipation coefficients
    that are used in ASH simulations in Section
    \,\ref{sec:validation-with-ash}. In the lower panel, the Prandtl
    number of unity is labeled by the horizontal dashed black line.
     }
  \label{fig:nu_kappa}
\end{figure*}

In EULAG, the implicit
treatment of sub-grid scales results in an additional contribution to
the kinetic energy and entropy evolution equations that is \textit{a
  priori} unknown. In the context of 3D
homogeneous turbulence in a cartesian box, \citet{Domaradzki:2003cb}
showed that the implicit sub-grid scales treatment of MPDATA (the
numerical algorithm behind EULAG) mimics qualitatively an eddy
viscosity. We build here on this idea and attempt to match the unknown
additional terms of Equations (\ref{eq:evol_KE}) and (\ref{eq:evol_S}) to
an eddy viscosity and an eddy thermal dissipation.
Such dissipation terms can be written as

\begin{align}
  \label{eq:diff_KE}
   \mathcal{V}_L =& \iint \left( -\boldsymbol{\nabla}\cdot\boldsymbol{\mathcal{D}}\right)_L
                     \cdot \mathbf{u}_L\, {\rm d}\Omega\, ,\\
  \label{eq:diff_S}
  \mathcal{K}_L =& \iint \left( Q_\kappa \right)_L \cdot S_L\, {\rm
                   d}\Omega\, ,
\end{align}

where $\boldsymbol{\mathcal{D}}$ and $ Q_\kappa$ respectively depend on an eddy viscosity $\nu$ and an eddy thermal
dissipation coefficient $\kappa$ (Equations
\ref{eq:visous_stress_tensor} and \ref{eq:heat_dissipation}). \reff{Using
the results from EULAG simulation, we calculate Equations (\ref{eq:diff_KE}) and
(\ref{eq:diff_S}) neglecting at this point the dependency upon radius
and scale $L$ of the unknown dissipation coefficients
$\nu$ and $\kappa$ (set to arbitrary values $\nu_a=1$, $\kappa_a=1$). Our
procedure to obtain effective dissipation coefficient is as follows. 
We average in time the right hand side of Equation
(\ref{eq:evol_KE}) (the left hand side is
vanishingly small because the system has reached a statistical
steady-state), and divide it with the analogously averaged Equation
(\ref{eq:diff_KE}). The resulting ratio gives us naturally $\nu_{\rm
  eff}$. The exact same procedure is applied to Equations
(\ref{eq:evol_S}) and (\ref{eq:diff_S}) to obtain $\kappa_{\rm eff}$.
We display the resulting} eddy-diffusion coefficients $\nu_{\rm eff}(r,L)$ 
and $\kappa_{\rm eff}(r,L)$ in the left panels of
Figure \ref{fig:nu_kappa} for case E1. 

Our numerical procedure embeds numerous numerical approximations that
need to be  acknowledged. First, we have no formal proof that the numerical
sub-grid scales effects can be matched to an eddy-type dissipation in
the case of turbulent convection simulations with EULAG. Second, EULAG
is formulated on a regular cartesian grid which is mapped to the
spherical geometry. Here, we carry our analysis on a spherical
harmonics decomposition that is obtained through an interpolation of
EULAG's results on Gauss-Legendre collocation points in latitude. Hence, some
numerical errors can appear in our analysis since this does not
reflect directly how Equations (\ref{eq:LBR})-(\ref{eq:LBRE_S}) are
solved in EULAG. Third, the spherical harmonics decomposition is
cut at the maximum accessible spherical harmonics degree $L_{\rm
  max}$, corresponding to the smallest grid size in latitude in
EULAG. We repeated our analysis by de-aliasing the spherical harmonics
decomposition up to $2L_{\rm max}/3$, which did not significantly change our
results, \reff{giving us confidence that the second and third error
  sources are not significantly affecting our study.}
\reff{Finally, because we cannot expect the sub-grid scale model to
  behave entirely as an eddy-type dissipation operator,
we consider a criterion to assess the robustness of the
effective eddy-dissipation coefficients we obtained.} For each value of
$\nu_{\rm eff}(r,L)$ (and $\kappa_{\rm eff}(r,L)$), we calculate the
standard deviation along the time window over which we averaged the various
contributions shown in Figure \ref{fig:nu_kappa_middle_cz}. If the
standard deviation is of the order of the deduced eddy-dissipation
coefficient, we do not confidently rely on its value. In the left
panels of Figure \ref{fig:nu_kappa} we darken the
pixels of the $(r,L)$ colormaps accordingly, and only consider the bright
pixels in the following analysis.

The robust eddy-diffusion coefficients show
characteristic patterns in $r$ and $L$, which are shown in the right
panels of Figure \ref{fig:nu_kappa}. The eddy viscosity is maximized
near the radial boundaries (which are impenetrable), and increases with $L$.
Its average value
lies between $5 \times 10^{11}$ cm$^2$/s and $1.5 \times 10^{12}$
cm$^2$/s which
agrees with empirically deduced implicit eddy-diffusion
coefficients obtained with mean-field models representing
EULAG simulations \citep[see, \textit{e.g.},][]{Simard:2016tk}. The average profile in $r$ and $L$
are shown in blue, and the standard deviation along $L$ and $r$ is
denoted by the gray areas. 

The eddy heat dissipation coefficient shows a reversed profile. It is
maximized at the center of the convection zone and decreases strongly
(by almost a factor of 8) close to the radial boundaries. In the robust
part of the $L$ profile ($L \ge 20$) it decreases with $L$.

The lower panels show the effective Prandtl number ${\rm Pr}_{\rm eff}
= \nu_{\rm eff}/\kappa_{\rm eff}$. It is generally assumed
in ILES that the Prandtl
number is close to $1$. Here we find that it is indeed very close to
unity in the middle of the convection zone (${\rm
  Pr}_{\rm eff}=1$ corresponds to white in the colormap), but rapidly increases by
almost an order of magnitude near the radial boundaries. The Prandtl
number also slightly increases with $L$, due to the simultaneous
increase of $\nu_{\rm eff}$ and decrease of $\kappa_{\rm eff}$.

We performed the same analysis for case E2, which is not shown
here. The radial profile of the dissipation coefficients is
very similar to the one shown in Figure \ref{fig:nu_kappa}, and the
$L$ profile is simply shifted to higher $L$'s thanks to the finer
resolution. The average value of
$\nu_{\rm eff}$ and $\kappa_{\rm eff}$ is decreased by factor between
$3$ and $5$, which is compatible with the naive expectation of doubling
the overall resolution (in both time and space) of the simulation,
leading to an implicit dissipation reduced by a factor of 4.

\subsection{Comparison with the ASH code}
\label{sec:validation-with-ash}

In order to further validate the deduced dissipation coefficients, we now use
them in LES done with the ASH code. The ASH code allows to specify
dissipation coefficient depending on both $r$ and $L$, which enables to
fully take into account the dissipation coefficients inverted in
Section \,\ref{sec:spectr-analys-result}.

We fit the dissipation coefficients with the following formulation
\begin{align}
  \nu_{\rm fit}(r,L) =&\nu_m\left(b_\nu L +
  c_\nu\right) \sum_{k=0}^{N} a^\nu_k
  \left(\frac{r}{R_\star}\right)^k \label{eq:fit_nueff}
\\
  \kappa_{\rm fit}(r,L) =&  \kappa_m \left(b_\kappa L +
  c_\kappa\right) \sum_{k=0}^{N} a^\kappa_k
  \left(\frac{r}{R_\star}\right)^k \, .
  \label{eq:fit_kaeff}
\end{align}

The radial shape of the effective dissipation coefficients is fitted
with a standard polynomial. 
The dependency against $L$ is fitted only in the $L\in[20,L_{\rm
  max}]$ range because the matched
$\nu_{\rm eff}$ and $\kappa_{\rm eff}$ are not statistically
significant at low $L$ (see the
discussion in previous section). We linearly fit both of them
for the sake of simplicity. 

At large scales, dissipation is small in EULAG and our method is not
able to match EULAG's dissipation to standard enhanced dissipation
coefficients. In ASH, we hence arbitrarily set them
to the small values $\nu_{l}=2\times 10^{11}$ cm$^2$/s and
$\kappa_{l}=5\times 10^{10}$ cm$^2$/s
using a smooth hyperbolic tangent (red curves in Figure
\ref{fig:nu_kappa}). Furthermore, an ASH simulation using these fitted
coefficients
is not stable when using the same grid resolution as in EULAG. Hence,
we double the resolution of case E1 in ASH and arbitrarily increase the value of
the fitted coefficients to $\nu_{s}= 10^{13}$ cm$^2$/s and
$\kappa_{s}=2\times 10^{13}$ cm$^2$/s at small scales (large $L$),
using again a smooth hyperbolic tangent.
\reff{The fitted profiles used in ASH are shown in red in the middle and right panels of Figure
\ref{fig:nu_kappa}, and can be written as (here, for the viscosity)
\begin{equation}
  \nu(r,L) = \left[ \nu_{\rm fit} + (1-s_l)\nu_l
               \right]\left(1-s_s\right) + 
                                                          s_s\nu_s\, ,
  \label{eq:full_func_form_dcoeff}
\end{equation}
where the subscripts $s$ and $l$ respectively refer to the small and
large scale branches of the profile, and the generic step function $s(L,L_i)$ is defined by
\begin{equation}
  \label{eq:step_function}
  s(L,L_i) =
  \frac{1}{2}\left[1+\tanh\left(\frac{L-L_i}{0.1\,L_i}\right)\right]\, ,
\end{equation}
where $L_i$ is centering parameter of the hyperbolic tangent.}

\begin{figure*}[htbp]
  \centering
  \hfill
     \includegraphics[width=0.95\linewidth]{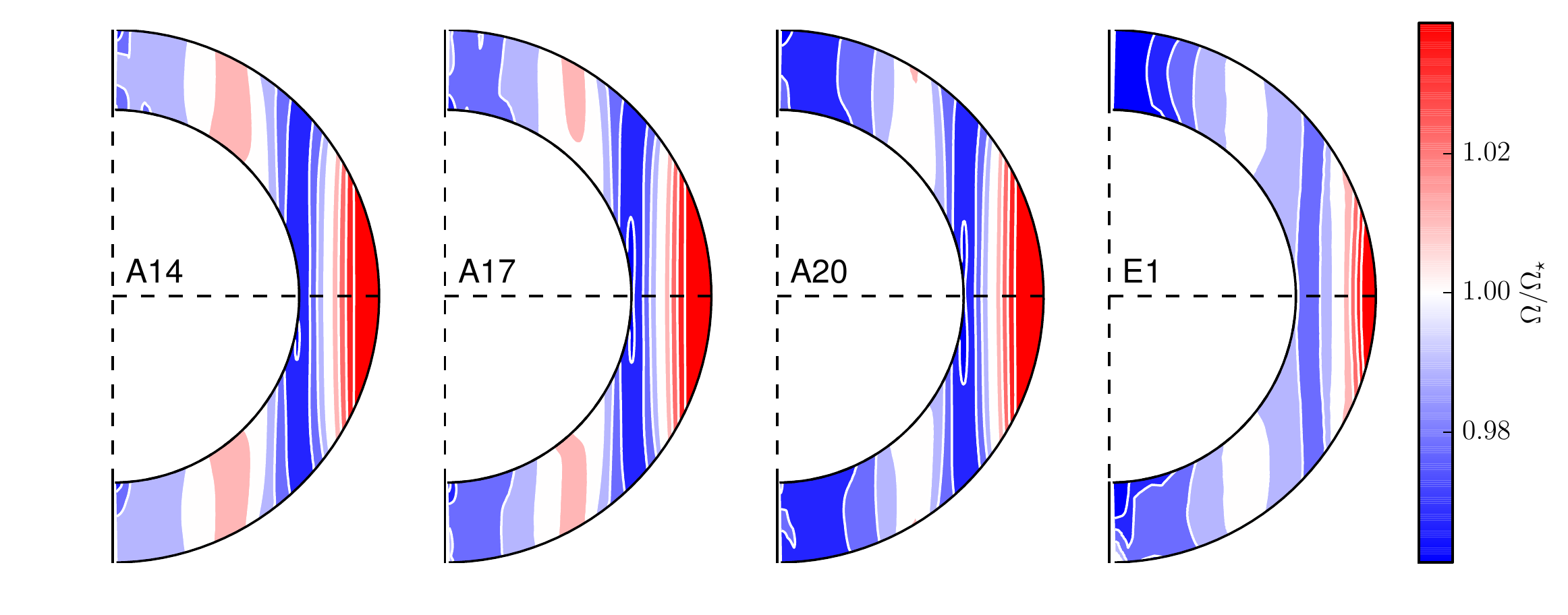} 
     \includegraphics[width=\linewidth]{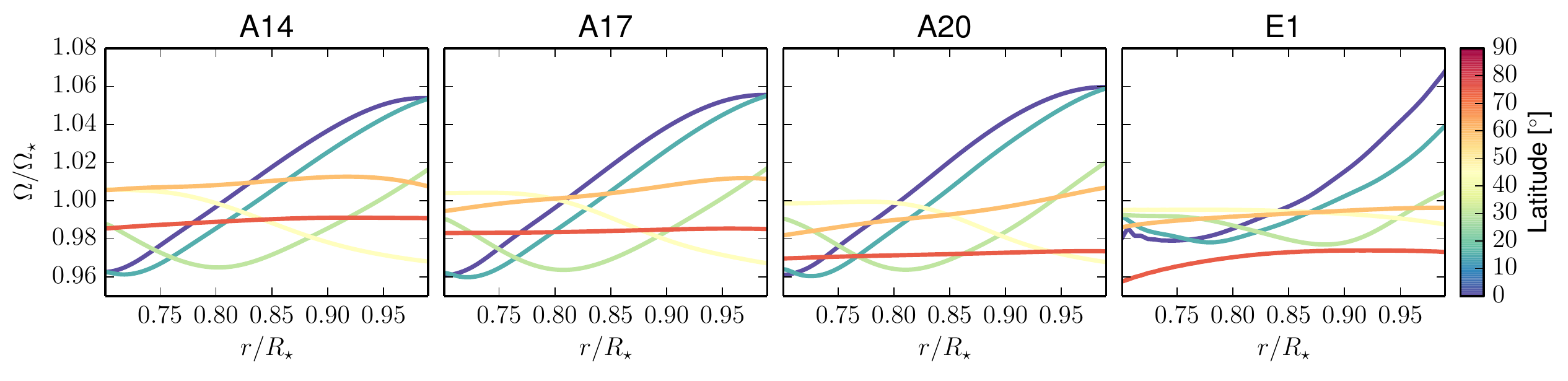} \\
     \hfill
     \includegraphics[width=0.95\linewidth]{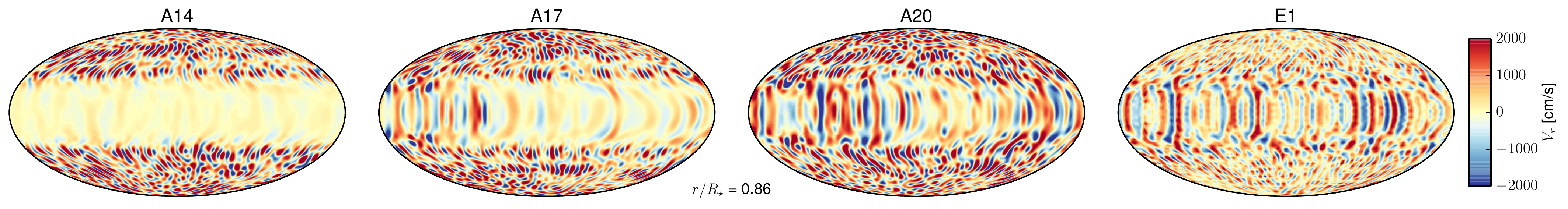} 
  \caption{\textit{Upper panels} Differential rotation for the three
    ASH cases (see red curves in Figure \ref{fig:nu_kappa}) and case E1. As in Figure \ref{fig:DR_Vr_S_Eulag}, blue
    and red respectively denote slower and faster rotation than the
    stellar rotation rate shown in white. \reff{\textit{Middle panels}
    Radial profile of the differential rotation at five different
    latitudes in the northern hemisphere. The profile at the equator
    is in blue, and at $75^\circ$ latitude in red (see color bar).} \textit{Lower panels} Radial
  velocity colormap at the middle of the convection zone for the ASH
  cases and case E1. Again, red denotes outward motions and
  blue downward motions.}
  \label{fig:compare_ASH_EULAG}
\end{figure*}

The ASH results are sensitive to the arbitrary choices of effective dissipation
coefficients at large and small scales. In order to illustrate their
sensitivity, we show three different cases obtained with ASH when the
transition at large scales is slightly shifted (see the solid, dashed and
dotted red lines in Figure \ref{fig:nu_kappa}). \reff{We label the cases
'A$L_l$' where $L_l$ is the centering scale of the hyperbolic tangent
at large scales. At small scales, the centering
scale is set to $L_s=64$ for all cases.} 
\reff{Figure \ref{fig:compare_ASH_EULAG} shows the resulting differential
rotation, its radial profiles at various latitudes, and the radial
velocity at mid depth, correspondingly for the three cases together
with the case E1.}

The three cases exhibit a differential rotation pattern which
qualitatively agrees with case E1. \reff{Overall, case A20 matches E1
  best, albeit its differential rotation is stronger, and steeper at
  the equator (blue line). However, A20 better matches E1 at mid
  latitudes (yellow to red lines) where A14 and A17 evince too strong
  differential rotation.} 

The convective patterns (lower
panels) change significantly between the three cases. In case A14,
convection is very weak on an equatorial band and is concentrated at
higher latitudes. In case A17 only an ``active nest of convection''
\citep[see][]{Brown:2008ii} is
observed on the equator, and in case A20 the convective patterns
qualitatively match E1 with slightly stronger convective
amplitudes. We recall that the only difference between cases A14 and
A20 is the location of the transition to low dissipation coefficients
at large scales, \textit{i.e.} in case A20 a larger range of large
scales is evolved with low dissipation coefficients. We recall that the kinetic
energy spectrum peaks in between $L=20$ and $L=30$ in the ASH
cases. The non-linear balance saturating the peak of the turbulent convective
spectrum is thus altered from left to right in Figure
\ref{fig:compare_ASH_EULAG}, with higher viscous and heat
dissipation in case A14 than in case A20 at the peak of the turbulent
spectrum. In case A14, the radial shear of the differential rotation near the equatorial plane is
strong enough to weaken significantly convection, whereas in case A20
the banana cells still live on.

\reff{In spite of these differences, the three ASH cases exhibit a
  similar convective heat transport (not shown) peaked at the
  center of the convection zone with an equivalent convective
  luminosity $L_{\rm cz}\sim 0.14\, L_\odot$ in cases A17 and A20, and
  $L_{\rm cz}\sim 0.13\, L_\odot$ in case A14.  All three are close to
  the E1 case, for which the convective transport reaches $0.12
  L_\odot$ (see Section \ref{sec:fiduc-stell-conv}).}

\begin{figure}[htb]
  \centering
  \hfill
     \includegraphics[width=0.95\linewidth]{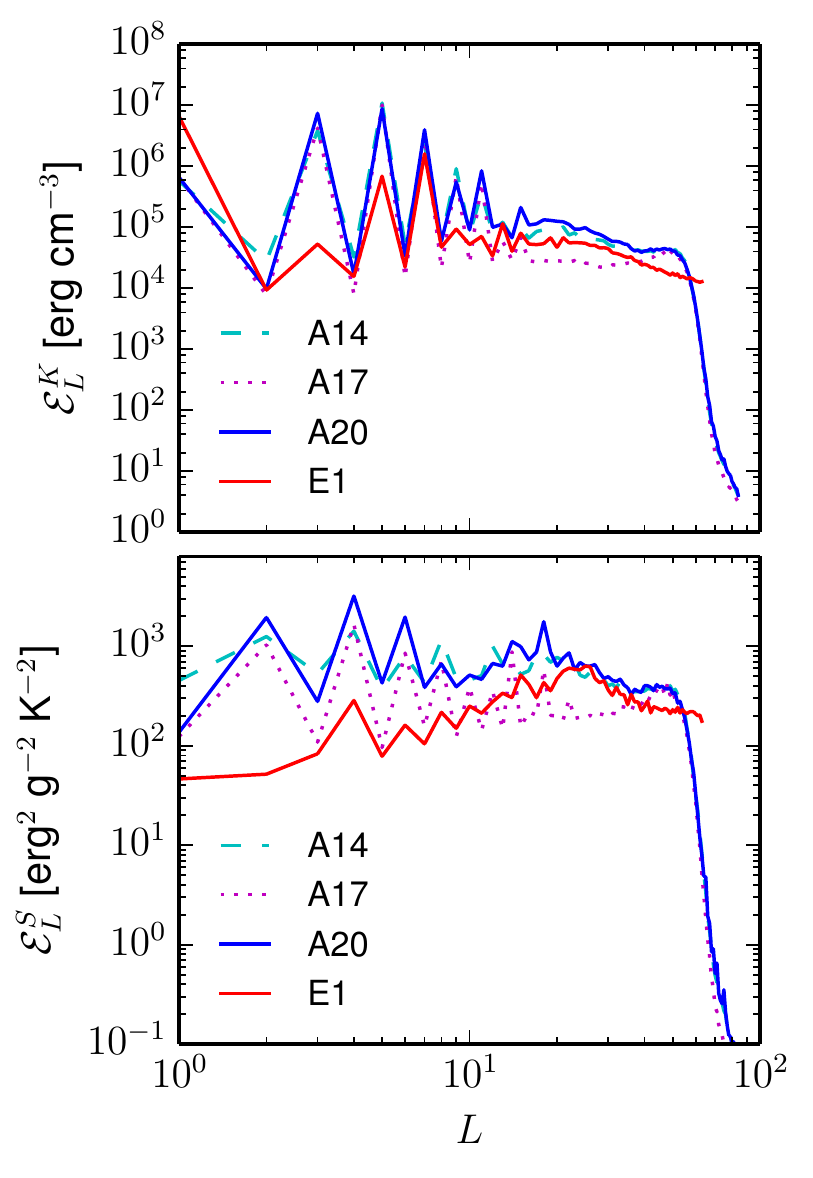} 
  \caption{\reff{Total kinetic energy (top panel) and entropy (bottom panel)
  spectra for cases A14 (dashed cyan), A17 (dotted magenta), A20
  (plain blue) and E1 (plain red).}}
  \label{fig:compare_ASH_EULAG_sp}
\end{figure}

\reff{Interestingly, the large-scale non-axisymmetric
entropy spectrum \reff{(Figure \ref{fig:compare_ASH_EULAG_sp})} is
significantly larger in ASH cases compared to EULAG,
while the kinetic energy spectrum at those scales remain the same. The
slope of the kinetic energy spectrum at mid-scale in EULAG is well recovered by
case A20, with a departure near the transition to high dissipation
around $L=64$.
The change in the spectra from case A14 to case A20 denotes a
non-trivial influence of scale-dependent dissipation
coefficients on the correlations between the large-scale heat and
momentum fluctuation. This may have important theoretical implications
for the amplitude of the large-scale convective fluctuations in the Sun
\citep{Hanasoge:2012fr,Greer:2015ff,Lord:2014hu}, and will be studied
in near future.}

\refff{Finally, we recall that the viscous heating has been omitted in
  the governing equations, as in ILES it is generally not accounted
  for by numerics supplying small-scale diffusion, and hence should
  not be formally introduced in this ILES-LES
  comparison. Nevertheless, omitting
  the viscous heating could formally lead to net energy loss
  in simulations, we hence intend to
  explore this effect in a subsequent study. 
}

Numerous numerical approximations have been made in \textit{(i)}
projecting the results from EULAG in spectral space, \textit{(ii)}
matching the imbalance in Equations (\ref{eq:evol_KE}) and
(\ref{eq:evol_S}) to standard laplacian dissipation terms, and
\textit{(iii)} using the matched dissipations coefficients in a LES
with the ASH code, introducing
arbitrary dissipation coefficients at
very large and very small scales. In spite of these approximations, we
managed to simulate with ASH (case A20), using fitted explicit dissipation
coefficients, a convective state producing a large scale differential
rotation that compares adequately with the results obtained with EULAG. This
comparison gives us confidence in the methodology we developed to estimate
the dissipation properties of ILES with EULAG, and \textit{a
  posteriori} \reff{suggests} that, at least in the bulk part of
the turbulent spectrum, EULAG's implicit dissipation \reff{could} be
interpreted in terms of standard explicit dissipation.

\section{Conclusions}
\label{sec:conclusions}

In this work we have studied the dissipation properties of implicit
large-eddy simulations of an idealized turbulent stellar convection
zone under the influence of rotation with the EULAG code. By
considering twin simulations where the grid resolution was doubled, we
showed that the kinetic energy spectrum of turbulent convection peaks
at smaller scales in the most refined model, as
expected. The latter model develops a stronger differential rotation,
which results from the complex interplay between the turbulent scales,
which are themselves affected by the implicit dissipation of EULAG.

In order to characterize this implicit dissipation, we developed a
spectral method to \textit{a posteriori} quantify the effect of the
sub-grid scale modelling of EULAG. By evaluating balance
equations for kinetic energy and entropy when the system has
reached a steady-state, we were able to isolate the implicit
dissipation contribution. The sub-grid scale modelling was shown to
match quantitatively well a standard laplacian-like operator from
medium to small scales. At large scales, the implicit dissipation
introduced by EULAG is very weak and could not be matched to such
classical formulation.

For a grid resolution comparable to previously published results with
EULAG in the context of solar dynamo
\citep{Ghizaru:2010im,Racine:2011gh,Beaudoin:2013eq,Passos:2014kx,Lawson:2015fq},
we find the effective viscosity and thermal diffusivities are of the
order of $10^{12}$ cm$^2$/s. However, we recall that the effective
dissipation does not match to an equivalent enhanced dissipation
coefficient for the large scales feature such as differential
rotation, for which we showed the effective dissipation to be 
a least an order of magnitude lower.

\reff{In order to further test our estimates of the dissipation
coefficients, we used them in a series of LES with the ASH
code. Note however that we were compelled to select
arbitrary dissipation coefficients at very small scales (below
the EULAG grid scale) and very large scales (where the implicit
dissipation in EULAG is small and does not match a standard laplacian
operator). We showed
that our arbitrary choice of dissipation coefficients at large
scale could significantly affect the large-scale flows in ASH. However, by
choosing adequately these dissipation coefficients, we were able to
reproduce the large scale
differential rotation of EULAG simulation. Our results thus indicate
that results from ILES and LES could be reconciled.}

The spectral analysis developed in this paper
is generic and can be easily extended
to the MHD regime \citep{Strugarek:2013kt}. A natural extension of
this work is to perform such joint simulations of cyclic dynamos, in
order to isolate to what extent the particular treatment of sub-grid
scales chosen in EULAG simulations impacts the existence and
characteristics of such solutions, as well as their dependency to
rotation. We intend to explore this aspect in a future publication.

\section*{Acknowledgments}

\reff{Constructive comments from two anonymous referees helped to
  improve the presentation. The authors
  thank J.F. Cossette for numerous discussions on convection
  modelling with EULAG, and N. Wedi and S. Malardel for discussions
  about the projection of the Coriolis force in spectral energy budgets.}
A. Strugarek is a National Postdoctoral Fellow at the Canadian
Institute of Theoretical Astrophysics. The authors acknowledge support from
Canada's Natural Sciences and Engineering Research Council. P. K. Smolarkiewicz is
supported by funding received from the European Research Council under
the European Union's Seventh Framework Programme (FP7/2012/ERC Grant
agreement no. 320375). This work
was also supported by the INSU/PNST, the ANR 2011 Blanc Toupies, and the ERC grant
STARS2 207430. S. Mathis acknowledges funding by the European Research
Council through ERC grant SPIRE 647383. We acknowledge access to
supercomputers through GENCI (project
1623), Prace (8th call), and ComputeCanada infrastructures.

\section*{References}

\bibliographystyle{elsarticle-harv}

\appendix

\section{Spectral transfers equations}
\label{sec:spectr-transf-equat}

\subsection{Vectorial spherical harmonics basis}
\label{sec:leftrlm-slm-tlmright}

\subsubsection{The standard vectorial basis}
\label{sec:definitions}

We define from \citet{Rieutord:1987go,Mathis:2005kz}:
\begin{align}
 \label{eq:RST_annex}
 \left\{
 \begin{array}{lcl}
   \Rlm{} &=& \Ylm{} \er \\
   \Slm{} &=& \dth\Ylm{}\etheta + \frac{1}{\sin{\theta}}\dphi\Ylm{}\ephi\\
   \Tlm{} &=& \frac{1}{\sin{\theta}}\dphi\Ylm{}\etheta  -\dth\Ylm{}\ephi
 \end{array}
 \right. ,
\end{align}
where $(\mathbf{e}_r,\mathbf{e}_\theta,\mathbf{e}_\varphi)$ defines the
spherical basis and $Y^m_l$ are the spherical harmonics
defined by
\begin{align}
 \label{eq:norm}
 Y_l^m =
 \sqrt{\frac{(2l+1)}{4\pi}\frac{(l-|m|)!}{(l+|m|)!}} P^{|m|}_l(\cos
 \theta)e^{im\varphi}\, 
\end{align}
where $P^m_l$ are the associated Legendre polynomials.
The basis \eqref{eq:RST_annex} have the following properties :
\begin{align}
\label{eq:R.R}
\iint \Rlm{1} \cdot \left( \Rlm{2} \right)^{cc} \dint{\Omega}{} =&
\delta_{l_1,l_2}\delta_{m_1,m_2} , \\
\iint \Slm{1} \cdot \left( \Slm{2} \right)^{cc} \dint{\Omega}{} =&
\iint \Tlm{1} \cdot \left( \Tlm{2} \right)^{cc} \dint{\Omega}{} 
                                                                     \nonumber
\\ =&
l_1(l_1+1)\delta_{l_1,l_2}\delta_{m_1,m_2} ,
\label{eq:S.S_T.T}
\end{align}
where $\dint{\Omega}{}=\sin\theta \mbox{d}\theta\mbox{d}\varphi$ the solid
angle, $cc$ means complex conjugate and $\delta$ is the Kronecker symbol. We also have:
\begin{align}
(\Slm{})^{cc} = (-1)^m\mathbf{S}^{-m}_l \, ,
\label{eq:conjg_rst}
\end{align}
and all the other scalar cross products are $0$. 

\subsubsection{Scalar fields identities}
\label{sec:scalar-fields}

Defining $\psi=\sumlm{}\left\{\psi_m^l(r)Y_l^m\right\}$, we get:
\begin{align}
 \label{eq:gradient}
 \grad\psi =& \sumlm{} \left\{ \dr\psi_m^l\Rlm{} +
   \frac{\psi_m^l}{r}\Slm{} \right\},\\
 \label{eq:lapla}
 \Div\grad\psi =& \sumlm{} \Delta_l\psi_m^lY_l^m
\end{align}
where $\Delta_l=\drr + \frac{2}{r}\dr-\frac{l(l+1)}{r^2}$. 

\subsubsection{Vectorial fields identities}
\label{sec:vectorial-fields}

For a vector 
\begin{align*}
\mathbf{X}=\sumlm{}\left\{\mathcal{A}^l_m\Rlm{} + \mathcal{B}^l_m\Slm{}
 + \mathcal{C}^l_m\Tlm{}\right\} \, ,  
\end{align*}
we obtain:
\begin{align}
  \label{eq:divergence}
  \Div\mathbf{X} =&
                    \sumlm{}\left[\frac{1}{r^2}\dr(r^2\MAlm{}) 
                    -l(l+1)\frac{\MBlm{}}{r}\right]Y_l^m\, , \\
  \label{eq:curl}
  \rot\mathbf{X} =&
                    \sumlm{}\left[l(l+1)\frac{\MClm{}}{r}\right]\Rlm{}\nonumber
  \\ 
                  &+\left[\frac{1}{r}\dr(r\,\MClm{})\right]\Slm{} \nonumber\\
                  &+\left[\frac{\MAlm{}}{r}-\frac{1}{r}\dr(r\, \MBlm{})\right]\Tlm{}\, , \\
  \nabla^2\mathbf{X} =&+\left[\Delta_l\MAlm{}-\frac{2}{r^2}(\MAlm{}-l(l+1)\MBlm{})\right]\Rlm{}
  \nonumber \\
 &+                  
\sumlm{}\left[\Delta_l\MBlm{}+2\frac{\MAlm{}}{r^2}\right]\Slm{}\nonumber\\
                  &  +
                    \left[\Delta_l\,\MClm{}\right]\Tlm{} \, .
                    \label{eq:lapla_vect}
\end{align}



\subsubsection{An alternative vectorial basis}
\label{sec:mathbfym_l-l+nu-basi}

The vectorial spherical harmonics basis defined in appendix
\ref{sec:leftrlm-slm-tlmright} is very efficient to calculate scalar
products or linear differential operator on vectors. Nevertheless, it
is quite hard to use it to express vectorial products. Instead we
define the following basis (\textit{e.g.}, see \citealt{Varshalovich:1988ul}): 
\begin{align}
 \label{eq:base2_annex}
 \Ylmn{} =  \sum_{\mu=-1}^{1}  (-1)^{l-m} c_{l,m,\nu,\mu}
 Y_{l+\nu}^{m-\mu} \mathbf{e}_\mu \, ,
\end{align}
where 
\begin{align}
 \label{eq:coeff_ylmn}
 c_{l,m,\nu,\mu} =\sqrt{2l+1}
 \left(
\begin{array}{ccc}
   l & l+\nu & 1 \\
   m & \mu-m & -\mu
 \end{array}
 \right) \, ,
\end{align}
$(\dots)$ is the $3$-j Wigner coefficient linked to Clebsch-Gordan
coefficients, and the vectors $\mathbf{e}_\mu$ are
\begin{align}
 \label{eq:emu}
 \left\{
 \begin{array}{lcl}
   \mathbf{e}_{-1} &=& \frac{1}{\sqrt{2}}\left(\mathbf{e}_x
     -i\mathbf{e}_y \right) \\
   \mathbf{e}_{0} &=& \mathbf{e}_z \\
   \mathbf{e}_{1} &=& - \frac{1}{\sqrt{2}}\left(\mathbf{e}_x
     +i\mathbf{e}_y \right)
 \end{array}
 \right. ,
\end{align}
where $(\mathbf{e}_x,\mathbf{e}_y,\mathbf{e}_z)$ defines
the cartesian basis.
Note that the equivalent of the conjugation rule \eqref{eq:conjg_rst}
is then 
\begin{align}
 \label{eq:conjg_Ylmn}
 \left(\mathbf{Y}^m_{l,l+\nu}\right)^{cc} =
 \left(-1\right)^{m+\delta_{0\nu}} \mathbf{Y}^{-m}_{l,l+\nu}\, .
\end{align}

\subsubsection{Vectorial product} 
\label{sec:vectorial-prodcut}

As previously, we decompose a vector $\mathbf{X}$ on this basis:
\begin{align*}
\mathbf{X} = \sumlmn{} X_{l,l+\nu}^{\hspace{0.1cm}m}\Ylmn{}.
\end{align*}
Evaluating the vectorial product of two vectors
$\mathbf{X}_{0} = \mathbf{X}_1\times \mathbf{X}_2$, one gets:
\begin{align}
 \Xlmn{0} = & \mysum{1}{2}{0}\nonumber \\ & \Xlmn{1}\Xlmn{2} 
 \mathcal{J}^{l_{0},m_{0},\nu_{0}}_{l_1,m_1,\nu_1,l_0,m_0,\nu_0} ,
 \label{eq:vect_prod}
\end{align}
where 
\begin{align}
 &\mathcal{J}^{l,m_1+m_2,\nu}_{l_1,m_1,\nu_1,l_2,m_2,\nu_2} =
 i(-1)^{\nu_1-\nu_2+(m_1+m_2)}\sqrt{\frac{3}{2\pi}}\nonumber
 \\
 &\sqrt{(2l_1+1)(2l_1+2\nu_1+1)(2l_2+1)(2l_2+2\nu_2+1)}\nonumber 
 \\
 &
   \sqrt{(2l+1)(2l+2\nu+1)}\left\{
\begin{array}{ccc}
   l_1 & l_2 & l \\
   l_1+\nu_1 & l_2+\nu_2 & l+\nu \\
   1 & 1 & 1
 \end{array}
 \right\}\nonumber \\ &
 \left(
\begin{array}{ccc}
   l_1 & l_2 & l \\
   m_1 & m_2 & -(m_1+m_2)
\end{array}
 \right) \nonumber \\ &
 \left(
\begin{array}{ccc}
  l_1+\nu_1 & l_2+\nu_2 & l+\nu  \\
  0 & 0 & 0
\end{array}
 \right),
 \label{eq:J_clebsch}
\end{align}
with $\{\cdots\}$ being the $9$-j Wigner coefficient.

\subsubsection{Scalar product} 
\label{sec:scalar-prodcut}

We decompose a vector $\mathbf{X}$ on this basis in the following way:
\begin{align*}
\mathbf{X} = \sumlmn{} X_{l,l+\nu}^{\hspace{0.1cm}m}(r)\Ylmn{}.
\end{align*}
Evaluating the scalar product of two vectors
$\mathcal{M} = \mathbf{X}_1\cdot \mathbf{X}_2$, one gets:
\begin{align}
 \label{eq:scal_prod}
 \mathcal{M}^l_m = \sum_{\substack{l_1,m_1,\nu_1 \\ l_2,m_2,\nu_2 \\ m_1+m_2 = m }}\Xlmn{1}\Xlmn{2} \mathcal{H}^{l,m_1+m_2}_{l_1,m_1,\nu_1,l_2,m_2,\nu_2}
\end{align}
where the sum symbol is the same as in Equation (\ref{eq:vect_prod}) and
\begin{align}
 &\mathcal{H}^{l,m_1+m_2}_{l_1,m_1,\nu_1,l_2,m_2,\nu_2} =
 (-1)^{l_1-(l_2+\nu_2)+l+m}\sqrt{\frac{1}{4\pi}}\nonumber
 \\
 &\sqrt{(2l_1+1)(2l_1+2\nu_1+1)}\nonumber\\ &\sqrt{(2l_2+1)(2l_2+2\nu_2+1)(2l+1)}\nonumber
 \\
 &
 \left\{
\begin{array}{ccc}
  l_1+\nu_1 & l_2+\nu_2 & l\\
   l_2 & l_1 & 1
 \end{array}
 \right\}\nonumber\\ &
 \left(
\begin{array}{ccc}
   l_1 & l_2 & l \\
   m_1 & m_2 & -(m_1+m_2)
\end{array}
 \right) \nonumber\\ &
 \left(
\begin{array}{ccc}
  l_1+\nu_1 & l_2+\nu_2 & l \\
  0 & 0 & 0
\end{array}
 \right)
 \label{eq:K}
\end{align}
with $\{\cdots\}$ being here the $6$-j Wigner coefficient. \\

\subsubsection{Basis change relations}
\label{sec:base-change-relat}

For a vector $\mathbf{X}$ decomposed in the following manner:
\begin{align*}
\mathbf{X} =& \sumlm{} \left\{\mathcal{A}^l_m\Rlm{} + \mathcal{B}^l_m\Slm{}
 + \mathcal{C}^l_m\Tlm{} \right\} \\
 =& \sumlmn{} \left\{X_{l,l+\nu}^{\hspace{0.1cm}m}\Ylmn{}\right\},
\end{align*}
we have the two following relations to change from one basis to the other:
\begin{align}
 &\left\{
   \begin{array}{lcl}
     \mathcal{A}^l_m &=& \frac{1}{\sqrt{2l+1}}\left[
       \sqrt{l}X_{l,l-1}^{\hspace{0.1cm}m} -
       \sqrt{l+1}X_{l,l+1}^{\hspace{0.1cm}m} \right] \\
    \mathcal{B}^l_m &=& \frac{1}{\sqrt{2l+1}}\left[
       \frac{1}{\sqrt{l}}X_{l,l-1}^{\hspace{0.1cm}m} +
       \frac{1}{\sqrt{l+1}}X_{l,l+1}^{\hspace{0.1cm}m} \right] \\
    \mathcal{C}^l_m &=& \frac{i}{\sqrt{l(l+1)}}X_{l,l}^{\hspace{0.1cm}m}
  \end{array}
 \right. \nonumber \, ,\\
 &\left\{
   \begin{array}{lcl}
     X_{l,l-1}^{\hspace{0.1cm}m} &=& \sqrt{\frac{l}{2l+1}}\left( \mathcal{A}^l_m +
       (l+1) \mathcal{B}^l_m\right) \\
     X_{l,l}^{\hspace{0.1cm}m} &=& -i\sqrt{l(l+1)} \mathcal{C}^l_m \\
     X_{l,l+1}^{\hspace{0.1cm}m} &=& \sqrt{\frac{l+1}{2l+1}}\left( - \mathcal{A}^l_m +
       l \mathcal{B}^l_m\right)
   \end{array}
 \right. .
 \label{eq:base_change}
\end{align}

\subsection{Kinetic Energy}
\label{sec:kinetic-energy}

We start from the momentum equation:
\begin{align}
 \partial_t \mathbf{u} &= -\left(\mathbf{u}\cdot \nabla
    \right)\mathbf{u} - 2\boldsymbol{\Omega}_\star\times\mathbf{u}
                         \nonumber \\
 &-\nabla \left(\frac{P}{\bar{\rho}}\right) - \frac{S}{c_p} \mathbf{g}  
 -
  \boldsymbol{\nabla}\cdot\boldsymbol{\mathcal{D}} 
         \label{eq:mom_conserv} 
\end{align}

We define the kinetic energy density spectrum by $\mathcal{E}^K_L = \frac{\bar{\rho}}{2}\iint
\mathbf{u}_L\cdot\mathbf{u}_L^{cc} \dint{\Omega}{}$. We multiply
Equation \eqref{eq:mom_conserv} by $\mathbf{u}_L$ and integrate it
over the spherical shell to obtain:
\begin{align}
 \dot{\mathcal{E}}^K_L(r) =& \mathcal{P}_L +
                              \mathcal{G}_L +
                              \mathcal{C}_{L} \nonumber \\ +& \sum_{L_1,L_2}
                              \mathcal{R}_L\left(L_1,L_2\right) + \mathcal{V}_L \, ,
 \label{eq:KE_evol}
\end{align}
where the various terms are given by
\begin{align}
 \label{eq:gradP_all}
 \mathcal{P}_L =& - \bar{\rho}\iint \nabla\left( \frac{P}{\bar{\rho}}  \right)_L
  \cdot \U_L^{cc} \, \dint{\Omega}{}\, , \\
 \label{eq:grav_all}
 \mathcal{G}_L =& - \bar{\rho}\iint \frac{S_L}{c_P}\mathbf{g} \cdot \U_L^{cc} \, \dint{\Omega}{}\, , \\
 \label{eq:cor_all}
 \mathcal{C}_L =& - \bar{\rho} \iint 2 \left(\boldsymbol{\Omega}_\star\times\mathbf{u}\right)_L\cdot\mathbf{u}_L^{cc} \, \dint{\Omega}{}\, , \\
 \label{eq:adv_all}
 \mathcal{R}_L =& - \bar{\rho} \iint
                   \left[\left(\mathbf{u}\cdot\nabla\right)\mathbf{u}\right]_L
                   \cdot \mathbf{u}_L^{cc} \, \dint{\Omega}{}\, , \\
 \label{eq:visc_all}
 \mathcal{V}_L =& \iint
                   \left(\boldsymbol{\nabla}\cdot\boldsymbol{\mathcal{D}}\right)_L
                   \cdot \mathbf{u}_L^{cc}\, \dint{\Omega}{}\, .
\end{align}

The pressure gradient contribution is easily calculated with Equation
(\ref{eq:gradient}) and the buoyancy contribution is also
straightforward to evaluate. We now detail the three remaining contributions.

\subsubsection{Advection}
\label{sec:advection-term}

We know that
\begin{align}
-\left(\U\cdot\bnab\right)\U= -\frac{1}{2}\bnab\left( \U\cdot\U\right)
   +\U\times\left(\bnab\times\U\right) 
\end{align}

As a result, one can simply use the standard formulae to evaluate the
scalar product (\ref{sec:scalar-prodcut}) and the
vectorial product (\ref{sec:vectorial-prodcut}) part of the
preceding equation to compute the full advection term \ref{eq:adv_all}.


\subsubsection{Coriolis force}
\label{sec:coriolis-term}

We recall the reader that the Coriolis force contribution vanishes for
the total energy. It is able though to redistribute energy spectrally
among scales, as we show now.
By definition we have 
\begin{align*}
 \mathbf{\Omega}_\star &= \Omega_\star\left( \cos\theta \mathbf{e}_r -
   \sin\theta\mathbf{e}_\theta \right) \\ &=
 \frac{\sqrt{4\pi}\Omega_\star}{\sqrt{3}}\left(\mathbf{R}_1^0 +
   \mathbf{S}_1^0\right) = \sqrt{4\pi}\Omega_\star \mathbf{Y}^0_{1,0}.
\end{align*}
Writing $\mathbf{X}=\mathbf{\Omega}_\star\times\mathbf{u}$, we note that:
\begin{align*}
 -\left(\mathbf{\Omega}_\star\times\mathbf{u} \right)^m_{l,l+\nu} &= \sqrt{4\pi}\sum_{\nu_1=-1}^{1}
   \left[ u^m_{l,l+\nu_1}\mathcal{J}^{l,m,\nu}_{l,m,\nu_1,1,0,-1}
                                                                    \right. \\
  +& 
     u^m_{l+1,l+1+\nu_1}\mathcal{J}^{l,m,\nu}_{l+1,m,\nu_1,1,0,-1} 
   \\
  +& \left.
   u^m_{l-1,l-1+\nu_1}\mathcal{J}^{l,m,\nu}_{l-1,m,\nu_1,1,0,-1}
     \right]\Omega_\star\, .
\end{align*}
The scalar product with $\U_L^{cc}$ is then trivial to compute. This
Coriolis contribution is equivalent to Equations (24-25) in
\citet{Augier:2013dz}, where a spectral kinetic energy equation is derived
in the context of General Circulation Models.





\subsubsection{Viscous tensor}
\label{sec:diffusion-term}

From the definition 
\begin{align*}
\boldsymbol{\mathcal{D}} = -2\bar{\rho}\nu\left(
 \boldsymbol{\epsilon} - \boldsymbol{I}\frac{\nabla\cdot\mathbf{u}}{3} \right)\, ,
\end{align*}
and the decomposition
\begin{align*}
 \mathbf{u} = \sum_{l,m} \mathcal{A}_m^l \Rlm{} + \mathcal{B}_m^l
 \Slm{} +\mathcal{C}_m^l \Tlm{},
\end{align*}
one may rewrite the components of the symmetric tensor in the
following fashion:
\begin{align*}
 \mathcal{D}_{rr} =& -2\bar{\rho}\nu\sum_{l,m}\left\{ \partial_r
   \mathcal{A}_m^l Y_l^m \right\} +
 \frac{2\bar{\rho}\nu}{3}\bnab\cdot\U \\
 \mathcal{D}_{r\theta} =& -\bar{\rho}\nu\sum_{l,m} \left\{ \left( \partial_r
   \mathcal{B}_m^l - \frac{\mathcal{B}_m^l}{r} +
   \frac{\mathcal{A}_m^l}{r}\right)\partial_\theta Y_l^m \right.\\ +& \left.
 \left( \partial_r \mathcal{C}_m^l - \frac{\mathcal{C}_m^l}{r}
 \right) \frac{\partial_\varphi Y_l^m}{\sin\theta} \right\} \\
 \mathcal{D}_{r\varphi} =& -\bar{\rho}\nu\sum_{l,m} \left\{ \left(
     \frac{\mathcal{C}_m^l}{r} - \partial_r \mathcal{C}_m^l
   \right)\partial_\theta Y_l^m \right. \\ +& \left. \left( \partial_r
   \mathcal{B}_m^l - \frac{\mathcal{B}_m^l}{r} +
   \frac{\mathcal{A}_m^l}{r} \right) \frac{\partial_\varphi Y_l^m}{\sin\theta} \right\} \\
 \mathcal{D}_{\theta\theta} =& -2\bar{\rho}\nu\sum_{l,m} \left\{
     \frac{\mathcal{B}_m^l}{r}\partial^2_{\theta\theta} Y_l^m +
     \frac{\mathcal{A}_m^l}{r} Y_l^m \right. \\ +& \left.
     \frac{\mathcal{C}_m^l}{r\sin\theta}\partial^2_{\theta\varphi}Y_l^m
     - \frac{\mathcal{C}_m^l}{r\tan\theta}\frac{\partial_\varphi
       Y_l^m}{\sin\theta} \right\} \\ +&
   \frac{2\bar{\rho}\nu}{3}\bnab\cdot\U \\
   \mathcal{D}_{\theta\varphi} =& -\bar{\rho}\nu\sum_{l,m} \left\{ 
     \frac{\mathcal{C}_m^l}{r\tan\theta}\partial_\theta Y_l^m
     -\frac{\mathcal{C}_m^l}{r} \partial^2_{\theta\theta} Y_l^m 
                                  \right. \\ +& \left.
     \frac{\mathcal{C}_m^l}{r\sin^2\theta}\partial^2_{\varphi\varphi}Y_l^m
     +
     \frac{2\mathcal{B}_m^l}{r\sin\theta}\partial^2_{\theta\varphi}Y_l^m
     \right. \\ -& \left. \frac{2 \mathcal{B}_m^l}{r\tan\theta}\frac{\partial_\varphi Y_l^m}{\sin\theta}
   \right\} \\
 \mathcal{D}_{\varphi\varphi} =& -2\bar{\rho}\nu\sum_{l,m} \left\{
   \frac{\mathcal{B}_m^l}{\sin^2\theta}\partial^2_{\varphi\varphi}Y_l^m
   + \frac{\mathcal{B}_m^l}{r\tan\theta}\partial_{\theta}Y_l^m \right.
                                 \\-& \left.
   \frac{\mathcal{C}_m^l}{r\sin\theta} \partial^2_{\theta\varphi}Y_l^m
   + \frac{\mathcal{C}_m^l}{r\tan\theta}\frac{\partial_\varphi Y_l^m}{\sin\theta}
   \right\} \\+& \frac{2\bar{\rho}\nu}{3}\bnab\cdot\U \, .
\end{align*}
Using the formula for the divergence of a tensor in spherical
coordinates, it can be shown, after a long but straightforward calculation, that:
\begin{align*}
 - \boldsymbol{\nabla}\cdot \boldsymbol{\mathcal{D}} =& \sum_{l,m} \alpha_m^l \Rlm{} + \beta_m^l
                             \Slm{} + \gamma_m^l \Tlm{}\, ,
\end{align*}
where
\begin{align*}
\alpha_m^l =& \bar{\rho}\nu\left[\Delta_l\mathcal{A}_m^l
     +\frac{1}{3}\partial^2_{rr}\mathcal{A}_m^l +
     \frac{2}{3r}\partial_r\mathcal{A}_m^l -
     \frac{8}{3r^2}\mathcal{A}_m^l \right. \\ &+ \left.
     \frac{l(l+1)}{3r}\left(-\partial_r\mathcal{B}_m^l
       +\frac{7}{r}\mathcal{B}_m^l\right)\right]    \\
 &+
 \frac{2}{3}\partial_r\left(\bar{\rho}\nu\right)\left(
       2\partial_r\mathcal{A}_m^l - 2\frac{\mathcal{A}_m^l}{r} +
       \frac{l(l+1)}{r}\mathcal{B}_m^l \right)  
 \\
 \beta_m^l =& 
\bar{\rho}\nu \left[ \Delta_l \mathcal{B}_m^l  -
     \frac{l(l+1)}{3r^2}\mathcal{B}_m^l
     +\frac{1}{3r}\left(\partial_r \mathcal{A}_m^l +
       \frac{8}{r}\mathcal{A}_m^l  \right) \right] \\
   &+ \partial_r\left(\bar{\rho}\nu\right) \left(
     r\partial_r\frac{\mathcal{B}_m^l}{r} + \frac{\mathcal{A}_m^l
     }{r} \right) \\
\gamma_m^l=&  \left\{
     \bar{\rho}\nu\Delta_l \mathcal{C}_m^l
     + \partial_r\left(\bar{\rho}\nu\right)r\partial_r\frac{\mathcal{C}_m^l}{r}
  \right\} \, ,
\end{align*}
with $\Delta_l = \partial_{rr}^2 + \frac{2}{r}\partial_r
- \frac{l(l+1)}{r^2}$.

\subsection{Entropy equation}
\label{sec:entropy-equation}

We start from the entropy equation:
\begin{align}
\partial_t S =
-\left(\mathbf{v}\cdot\nabla\right) \left(S+S_{a}\right)
 -\frac{S}{\tau} + Q_{\kappa} \, ,
\label{eq:entropy_eq}
\end{align}
where we neglected viscous heating for the sake of simplicity. The
entropy spectrum is defined by $\mathcal{E}^S =
\frac{1}{2}\iint S_L^2\,{\rm d}\Omega$. We
multiply Equation (\ref{eq:entropy_eq}) by $S_L$ and integrate it over
the spherical shell to obtain:
\begin{align}
  \dot{\mathcal{E}}^S_L(r) = \mathcal{S}^a_L + \mathcal{N}_L +
  \sum_{L_1,L_2}\mathcal{A}_L\left(L_1,L_2\right) + \mathcal{K}_L
\, ,
  \label{eq:S_KE_appendix}
\end{align}
where the various terms are given by
\begin{align}
  \label{eq:ambient_contrib}
  \mathcal{S}^a_L =& - \iint
                     \left[\left(\mathbf{v}\cdot\nabla\right)
                     S_{a}\right]_L \cdot S_L^{cc}\, \dint{\Omega}{}\, , \\
  \mathcal{N}_L = & - \iint \left(\frac{S}{\tau}\right)_L \cdot
                    S_L^{cc}\, r^2\dint{\Omega}{}\, , \\
  \mathcal{A}_L = & - \iint
                     \left[\left(\mathbf{v}\cdot\nabla\right)
                     S \right]_L \cdot S_L^{cc}\, \dint{\Omega}{}\, , \\
  \mathcal{K}_L = & \iint \left(Q_\kappa\right)_L  \cdot
                    S_L^{cc}\, r^2\dint{\Omega}{}\, .
\end{align}

The non-linear contribution $\mathcal{A}_L$ is then calculated using
the scalar product (\ref{eq:scal_prod}), and the three other
contributions are easily calculated using the spherical harmonics
selection rules.

\end{document}